%% file: jswml.tex
\documentclass[fleqn,useAMS,usenatbib]{mn2e}
\usepackage{amsmath}
\usepackage{color}
\usepackage[pdftex]{graphicx}
\usepackage{times}
\usepackage{url}

\newcommand{\Cov}{\mathrm{Cov}}
\newcommand{\hMpc}{\mbox{$h^{-1}{\rm Mpc}$}}
\newcommand{\volunit}{\mbox{$h^{-3}{\rm Mpc}^{3}$}}
\newcommand{\denunit}{\mbox{$h^{3}{\rm Mpc}^{-3}$}}

\newcommand{\sbunit}{\mbox{mag arcsec$^{-2}$}}
\newcommand{\Sersic}{S\'{e}rsic}
\newcommand{\Var}{\mathrm{Var}}
\newcommand{\Vmax}{\mbox{$V_{\rm max}$}}
\newcommand{\Vdcmax}{\mbox{$V^{\rm dc}_{\rm max}$}}

\title[GAMA LF evolution]
{Galaxy and Mass Assembly (GAMA): maximum likelihood determination of the 
luminosity function and its evolution}

\input authors.tex

\begin{document}

\maketitle

\begin{abstract}
  We describe modifications to the joint stepwise maximum likelihood
  method of Cole (2011) in order to simultaneously fit the GAMA-II
  galaxy luminosity function (LF), corrected for radial density variations,
  and its evolution with redshift.  The whole sample is reasonably well-fit
  with luminosity ($Q_e$) and density ($P_e$) evolution parameters
  $Q_e, P_e \approx 1.0, 1.0$ 
  but with significant degeneracies characterized by
  $Q_e \approx 1.4 - 0.4 P_e$.
  Blue galaxies exhibit larger luminosity density evolution than red galaxies,
  as expected.
  We present the evolution-corrected $r$-band LF for the whole
  sample and for blue and red sub-samples, using both Petrosian and
  \Sersic\ magnitudes.
  Petrosian magnitudes miss a substantial fraction of the flux
  of de Vaucouleurs profile galaxies: the \Sersic\ LF is substantially higher 
  than the Petrosian LF at the bright end.
\end{abstract}


\section{Introduction} \label{sec:intro}

The luminosity function (LF) is perhaps the most fundamental model-independent
quantity that can be measured from a galaxy redshift survey.
Reproducing the observed LF is the first requirement of a successful
model of galaxy formation,
and thus accurate measurements of the LF are important in constraining
the physics of galaxy formation and evolution (e.g. \citealt{Benson2003}).
In addition, accurate knowledge of the survey selection function
(and hence LF) is required in order to determine the clustering of a
flux-limited sample of galaxies \citep{Cole2011}.

A standard $1/\Vmax$ \citep{Schmidt1968} estimate of the LF
is 
vulnerable
to radial density variations within the sample.
This 
vulnerability
can be largely mitigated by multiplying the maximum volume 
in which each galaxy is visible, $\Vmax$,  by the integrated radial overdensity 
of a density-defining population \citep{Baldry2006,Baldry2012}.
Maximum-likelihood methods \citep{Sandage1979,Efstathiou1988}, 
which assume that the luminosity and spatial dependence of the galaxy 
number density are separable, are, by construction, 
insensitive to density fluctuations.
However, if the sample covers a significant redshift range, 
galaxy properties (such as luminosity) and number density are subject
to systematic evolution with lookback time.
All of the above methods must then either be applied to restricted redshift
subsets of the data, or be modified to explicitly allow for evolution
(e.g. \citealt{Lin1999,Loveday2012}).

\citet{Cole2011} recently introduced a {\em joint stepwise maximum likelihood} 
(JSWML) method, which jointly fits non-parametric estimates of the LF 
and the galaxy overdensity in radial bins, along with an evolution model.
In this paper we describe modifications made to the JSWML method
in order to successfully apply it
to the Galaxy and Mass Assembly (GAMA) survey \citep{Driver2011}.
In the GAMA-II sample, $L^*$ galaxies can be seen out to redshift
$z \approx 0.35$, and so one has a reasonable redshift baseline over which
to constrain luminosity and density evolution.
\citet{Loveday2012} have previously investigated LF evolution in the
GAMA-I sample, finding that at higher redshifts:
all galaxy types were more luminous,
blue galaxies had a higher comoving number density
and red galaxies had a lower comoving number density.
Here we exploit the greater depth (0.4 mag) of GAMA-II versus GAMA-I,
and use an estimator of galaxy evolution that does not assume a parametric form
(e.g. a Schechter function) for the LF.

The paper is organized as follows.
In Section~\ref{sec:data} we describe the GAMA data used along with
corrections made for its small level of incompleteness.
Our adopted evolution model is described in Section~\ref{sec:ev_model}
and the density-corrected \Vmax\ method in Section~\ref{sec:Vmax}.
Methods for determining the evolution parameters are discussed in 
Section~\ref{sec:ev}.
We present tests of our methods using simulated data in Section~\ref{sec:tests}
and apply them to GAMA data in Section~\ref{sec:results}.
We briefly discuss our findings in Section~\ref{sec:discussion}
and conclude in Section~\ref{sec:concs}.

Throughout, we assume a Hubble constant of 
$H_0 = 100 h \mbox{ km s}^{-1} \mbox{ Mpc}^{-1}$ 
and an $\Omega_M = 0.3, \Omega_\Lambda = 0.7$ cosmology in
calculating distances, co-moving volumes and luminosities.

\section{GAMA-II data, $K$- and completeness corrections} \label{sec:data}

In April 2013 the GAMA survey completed spectroscopic coverage of the
three equatorial fields G09, G12 and G15.
In GAMA-II, these fields were extended in area to cover $12 \times 5$ degrees
each\footnote{
The RA, dec ranges of the three fields, all in degrees, are
G09: 129.0--141.0, $-2.0$--$+3.0$;
G12: 174.0--186.0, $-3.0$--$+2.0$;
G15: 211.5--223.5, $-2.0$--$+3.0$.
}
and all galaxies were targeted to a Galactic-extinction-corrected
SDSS DR7 Petrosian $r$-band magnitude limit of $r = 19.8$ mag.
In our analysis, we include all main-survey targets
({\sc survey\_class} $>= 4$)\footnote{Note that 
in this latest version of TilingCat, objects that failed visual inspection
({\sc vis\_class} = 2, 3 or 4) also have {\sc survey\_class} set to zero.} 
with reliable {\sc autoz} \citep{Baldry2014}
redshifts ($nQ \ge 3$) from TilingCatv43 \citep{Baldry2010}.
Redshifts (from DistancesFramesv12) are corrected for local flow using the 
\citet{Tonry2000} attractor model as described by \citet{Baldry2012}.

\begin{figure}
\includegraphics[width=\linewidth]{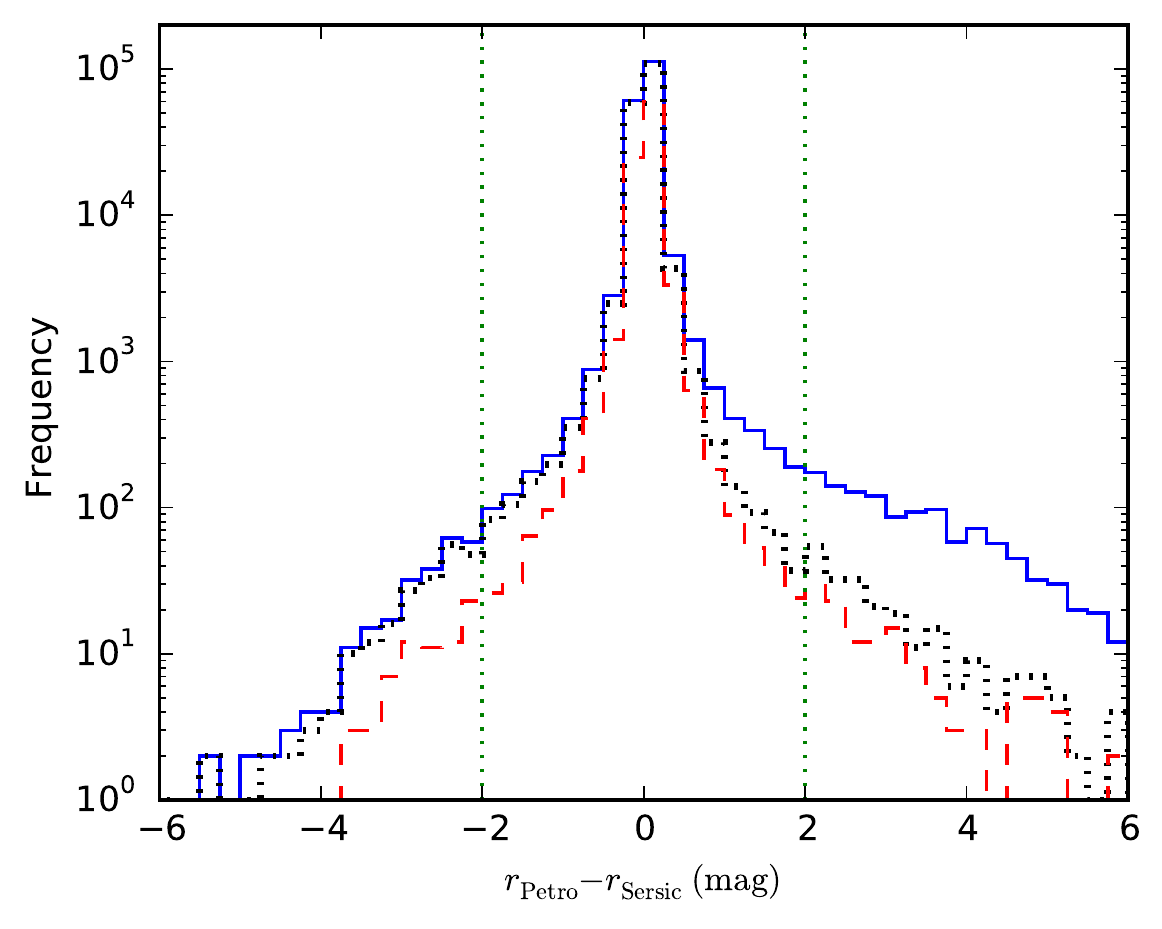}
\caption{
Histogram of the difference between Petrosian and \Sersic\ magnitudes
for all GAMA-II main-survey targets (continuous blue histogram),
and for targets without a nearby bright stellar neighbour, 
as defined in the text (black dotted histogram).
The dashed red histogram indicates the subset of the
latter targets classified as red.
The vertical dotted lines denote the additional constraint
$|r_{\rm Petro} - r_{\rm Sersic}| < 2.0$ mag
required for galaxies to be assumed uncontaminated; 
only about 0.3 per cent of remaining targets lie beyond these limits.
}
\label{fig:dmhist}
\end{figure}

We calculate LFs using both \citet{Petrosian1976} and \citet{Sersic1963}
photometry, corrected
for Galactic extinction using the dust maps of \citet{Schlegel1998}.
We use 
single
\Sersic\ model magnitudes truncated at ten effective radii
as fit by \citet{Kelvin2012}.
Kelvin et al. show that these recover essentially all of the
flux for an $n = 1$ (exponential) profile, and about 96 per cent of
the flux of an $n = 4$ (de Vaucouleurs) profile.
SDSS Petrosian magnitudes, while also measuring almost all of the flux
for exponential profiles, measure only about 82 per cent of the flux for
de Vaucouleurs profiles \citep{Blanton2001}.
\Sersic\ magnitudes are, however, more susceptible to contamination from nearby
bright objects, which can cause them to be overestimated by several mag.
We identify galaxies which may have contaminated photometry by searching
for brighter stellar neighbours within a distance, 
up to a maximum of five arcmin, of twice the star's
isophotal radius (\verb|isoA_r| in the SDSS \verb|PhotoObj| table).
Five per cent of GAMA targets are flagged in this way.

Fig.~\ref{fig:dmhist} shows a histogram of 
$\Delta m = r_{\rm Petro} - r_{\rm Sersic}$
for all GAMA-II main-survey targets (continuous blue histogram),
and for targets without a nearby bright stellar neighbour 
(black dotted histogram).
The majority (about 72 per cent) of excluded galaxies have positive
$\Delta m$, i.e. are brighter in \Sersic\ than Petrosian magnitude.
The dashed red histogram indicates targets without a nearby bright stellar 
neighbour that are classified as red 
(as defined towards the end of this section).
It is clear from this figure that uncontaminated
red galaxies preferentially have brighter 
\Sersic\ than Petrosian magnitudes.
This is as expected, assuming that they are bulge-dominated, 
and hence have profiles with higher \Sersic\ index.

We exclude an additional 487 targets (0.3 per cent of the total)
for which the $r$-band \Sersic\ and Petrosian magnitudes differ by 
more than 2 mag.
This magnitude difference cut is somewhat arbitrary, but
is designed to exclude galaxies with bright stellar neighbours that do not quite
satisfy the above criterion (for instance if a galaxy lies on a star's
diffraction spike) or with bad sky background determination.
It seems extremely unlikely that the \Sersic\ magnitude would recover more flux
than this from an uncontaminated galaxy.

\begin{table}
 \caption{Classification of the 487 GAMA targets without a bright 
stellar neighbour (as defined in the text) for which 
$\Delta m = r_{\rm Petro} - r_{\rm Sersic}$ lies outside the range [$-2$, 2] mag.
See text for meaning of first column.  
}
 \label{tab:outliers}
 \begin{tabular}{rrr}
 \hline
Class & \multicolumn{1}{c}{$\Delta m > 2$} & \multicolumn{1}{c}{$\Delta m < -2$} \\
 \hline
OK & 12 & 42\\
Deblend & 19 & 56\\
FSC & 13 & 22\\
BSC & 144 & 16\\
Merger & 56 & 37\\
Sky & 17 & 46\\
NO & 7 & 0\\
Total & 268 & 219\\
 \hline
 \end{tabular}
\end{table}

We have visually inspected these additional culled targets, for which 
$|\Delta m| = |r_{\rm Petro} - r_{\rm Sersic}| > 2$~mag, 
and placed them in one of the following categories:
OK: no obvious problem; 
Deblend: large galaxy image likely to have been shredded by
the SDSS deblending algorithm;
FSC: nearby faint stellar companion (comparable to or fainter than target);
BSC: nearby bright stellar companion (much brighter than target);
Merger: nearby galaxy companion(s);
Sky: bad sky background;
NO: no object visible.
The number of targets falling into each category, subdivided by whether
$\Delta m$ is positive (\Sersic\ flux is brighter) or negative 
(Petrosian flux is brighter) is given in Table~\ref{tab:outliers}.
For the former sample, just over half of the
cases of possibly overestimated \Sersic\ flux appear to be due to a nearby
star which has more successfully been excluded from the 
Petrosian flux estimate.
For the latter sample, the most common cause of underestimated \Sersic\ flux 
or overestimated Petrosian flux is likely due to deblending issues
or a bad sky determination.
We note that the presence of a nearby bright star should be totally
uncorrelated with a galaxy's intrinsic properties, and so excluding 
targets for this reason should not bias the sample in any way.
A small bias could be caused by excluding the $\simeq 20$ per cent of 
inspected galaxies (about 0.06 per cent of total targets)
for which the suspect photometry is caused by a neighbouring galaxy,
since galaxies in crowded regions are expected to be more luminous than average.
In cases when the Petrosian and \Sersic\ magnitudes differ by more than 2 mag,
both magnitude estimates are suspect, and so it is debatable whether these 
objects should be GAMA targets at all.
At worst, the effect of excluding targets with bright stellar neighbours
or discrepant magnitudes (5.3 per cent of the entire GAMA-II sample)
will be to bias the LF normalization low by up to five per cent.

After excluding GAMA main survey targets with either an unreliable redshift
(1.2 per cent) or suspect photometry (5.3 per cent), 
we are left with a sample of 173,527 galaxies in the redshift range
$0.002 < z < 0.65$.

To determine $K$-corrections, we use {\sc kcorrect} v4.2 
\citep{Blanton2007} to fit spectral energy distributions to
$ugriz$ GAMA matched-aperture SExtractor \citep{Bertin1996} {\sc AUTO} 
magnitudes taken from ApMatchedCatv04 \citep{Hill2011}.
As shown in Appendix B of \citet{Taylor2011} and Fig.~17 of
\citet{Kelvin2012}, SDSS model magnitudes,
which have been recommended for calculating galaxy colours,
e.g. \citet{Stoughton2002}, are ill-behaved for galaxies of intermediate
\Sersic\ index which are well fit by neither pure exponential nor pure 
de Vaucouleurs profiles.
GAMA matched-aperture magnitudes do not force a particular functional form
on the galaxy profile and so provide more reliable colours for all galaxy types.
In practice, we find that the choice of magnitude type used for $K$-corrections
makes little difference to our LF estimates, with the Schechter fit parameters
changing by less than 1-sigma.
We use $K$-corrections to reference redshift $z_0 = 0.1$ in order to allow 
direct comparison with previous results \citep{Loveday2012}.
For the three GAMA-II targets which are missing {\sc AUTO} magnitudes,
and for the 3.3 per cent of targets for which {\sc kcorrect} reports
a $\chi^2$ statistic of 10.0 or larger, implying a poor SED fit,
we set the $K$-correction to the mean of the remaining sample.
We have visually inspected 235 of these targets with poor-fitting SEDs.
About 29 per cent are close to a bright star or are otherwise likely to suffer
from poorly-estimated sky background; 
about 22 per cent have one or more close neighbours
and may thus suffer contaminated photometry;
about 13 per cent show evidence of AGN activity.
The remaining 35 per cent show no obvious reason for the SED fit to be poor,
but it seems likely that many of these cases may be due to poor $u$-band 
photometry with underestimated errors.

\begin{figure}
\includegraphics[width=\linewidth]{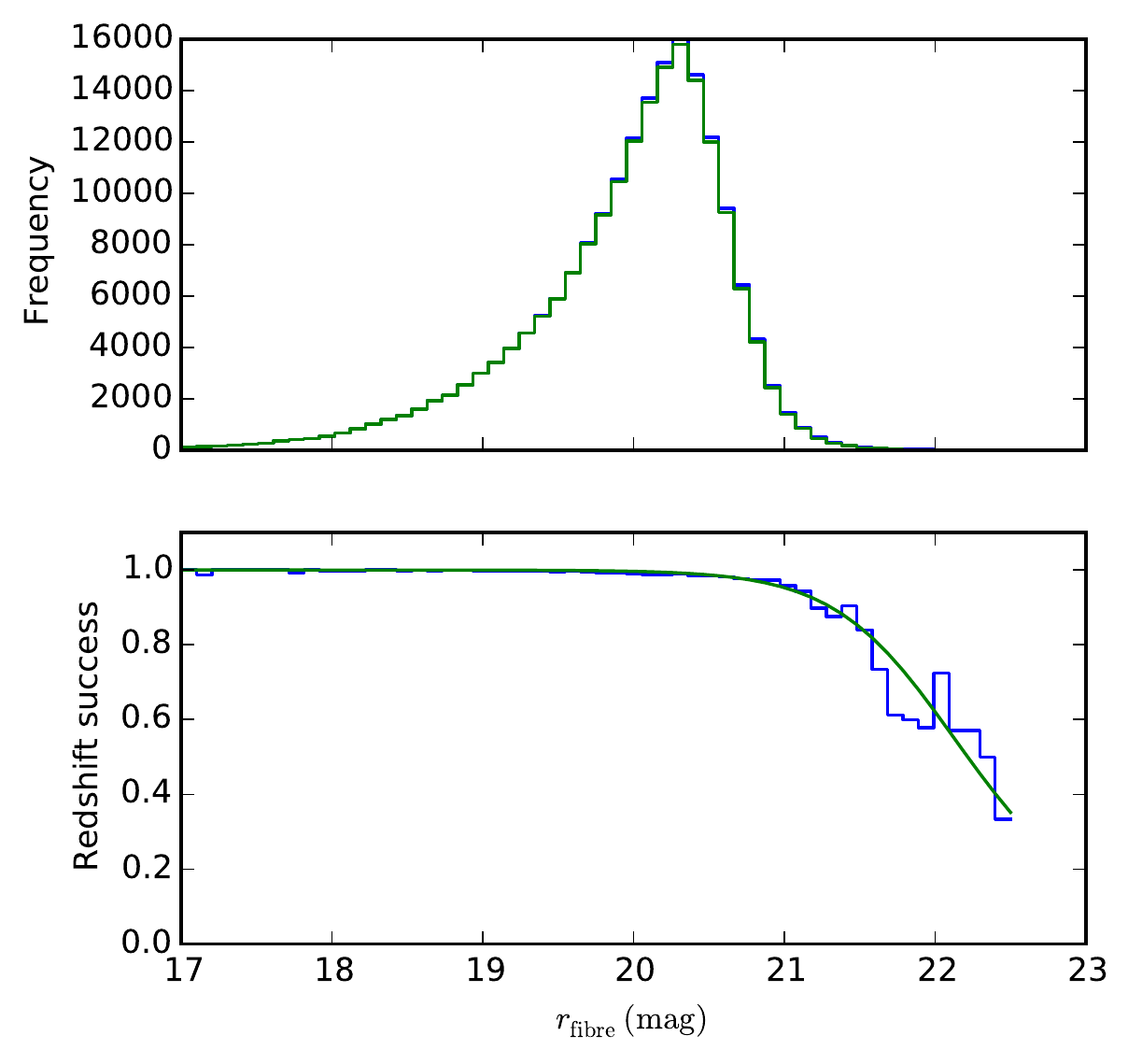}
\caption{Redshift success rate as a function of $r$-band fibre magnitude.
The top panel shows histograms of $r_{\rm fibre}$ for all observed
galaxies in blue, and galaxies with a reliable redshift measurement ($nQ > 2$)
in green.
Redshift success, the ratio of the latter to the former, is shown as a histogram
in the lower panel, along with a best-fit sigmoid-type function.
The large fluctuations at faint magnitudes ($r_{\rm fibre} > 21$)
are simply due to small-number statistics: 
the success rate is the ratio of two small numbers.
}
\label{fig:zcomp}
\end{figure}

While SDSS DR7 has improved photometric calibration over DR6 (used for selection
of GAMA-I targets), it will suffer from the same surface-brightness-dependent
selection effects as DR6, and so we assume the same {\em imaging completeness}
$C_{\rm im}$ as shown in Fig.~1 of \citet{Loveday2012}.
In this paper, we only measure the $r$-band LF, and so assume that
{\em target completeness} is 100 per cent.
In fact, just 0.1 per cent of GAMA-II main targets with $r < 19.8$ mag 
lack a measured spectrum, with no systematic dependence on magnitude
(Liske et al., submitted to MNRAS).
Since GAMA-II uses a new, fully-automated redshift measurement 
\citep{Baldry2014}, we have re-assessed {\em redshift success rate}
for GAMA-II.
Fig.~\ref{fig:zcomp} shows redshift success rate, defined as the fraction
of observed galaxies with reliable ($nQ \ge 3$) redshifts, 
as a function of $r$-band fibre magnitude.
This success rate is well-fit by a modified sigmoid function
\begin{equation}
  \label{eq:zcomp}
  C_z = [1 + e^{a(r_{\rm fibre} - b)}]^{-c}
\end{equation}
with parameters $a = 2.55$ mag$^{-1}$, $b = 22.42$ mag and $c = 2.24$.
The extra parameter $c$ (c.f. \citealt{Ellis2007,Loveday2012})
is introduced to provide a more extended decline
in $C_z$ around $r_{\rm fibre} \approx 20$ mag.
Without it, the sigmoid function drops too sharply to faithfully follow the 
observed $C_z$.

\begin{figure}
\includegraphics[width=\linewidth]{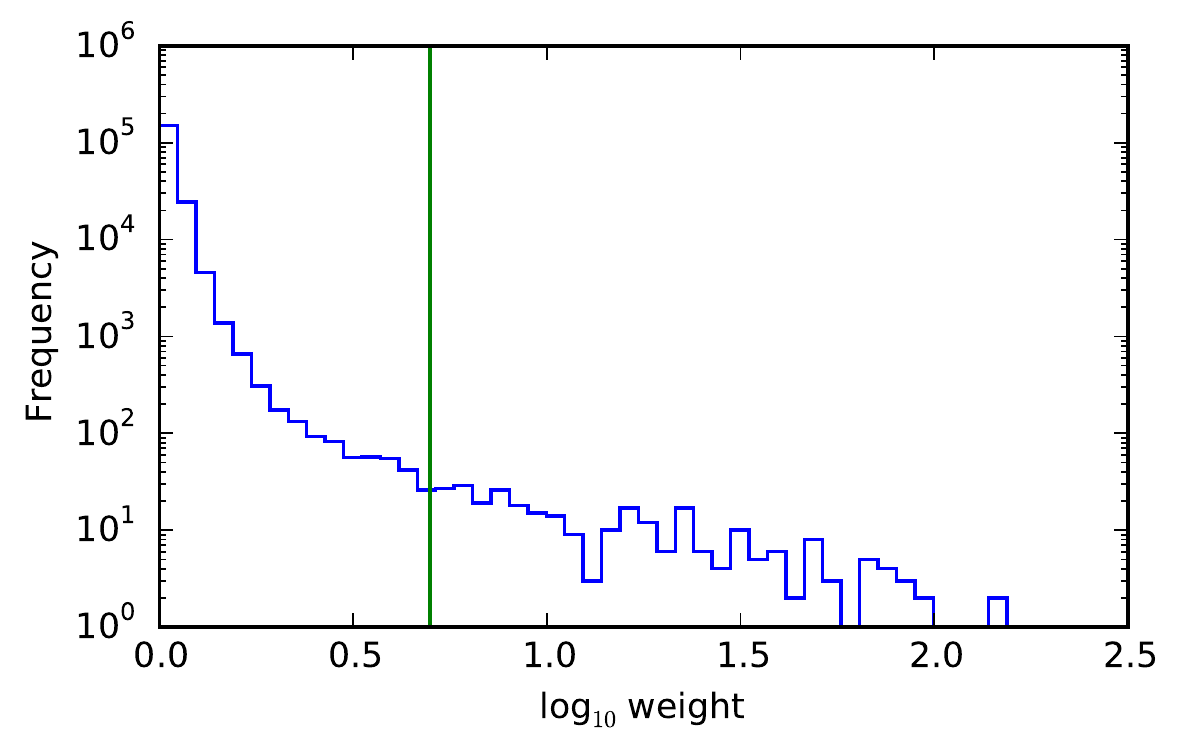}
\caption{Histogram of completeness-correction weights for GAMA-II galaxies.
Note that both axes use logarithmic binning.
The vertical line corresponds to the applied upper limit weight cap of 5.0.
}
\label{fig:weight}
\end{figure}

Each galaxy is given a weight equal to the reciprocal of the product
of imaging completeness and redshift success rate,
$W_i = 1/({C_{\rm im}}_i {C_z}_i)$.
A histogram of these weights is shown in Fig.~\ref{fig:weight}.
While the vast majority of galaxies (99.5 per cent) have $W_i < 2$,
there is a tail of rare objects with weights as high as 100 or more.
We have visually inspected the 157 objects with an assigned weight above 10.0.
Of these, 38 per cent are close to a bright star or are otherwise likely 
to have a poorly-determined sky background; 
another 38 per cent have nearby neighbouring galaxies
which might lead to a compromised surface-brightness estimate;
10 per cent are isolated and show no obvious visual indication of being of 
low surface-brightness.
That left just 14 per cent which appeared to be genuine low surface-brightness
galaxies, potentially with half-light surface brightness
$\mu_{50,r} \ga 24\ \sbunit$
and/or with fibre magnitude $r_{\rm fibre} \ga 22$ mag.
We therefore chose to set an upper limit cap of 5.0 on incompleteness weights,
i.e. to set $W_i = \min(W_i, 5.0)$.
This limit corresponds to the inverse redshift success rate for galaxies
with the faintest fibre magnitudes (Fig.~\ref{fig:zcomp}).
While only 297 galaxies (0.16 per cent of the total) have $W_i > 5.0$, 
these galaxies are likely to lie at the extreme faint end of the LF,
where there are few observed galaxies, and so spurious weights could
potentially bias the LF faint end.
The mean galaxy weights before and after applying this cap are 1.12 and
1.09, respectively.

The effect of applying this weight cap is to reduce the best-fit value
of the density evolution parameter $P_e$ by about 40 per cent,
with a corresponding increase in the best-fit value of the 
luminosity evolution parameter $Q_e$.
Best-fit LF parameters change by less than 1-$\sigma$.

When subdividing GAMA galaxies into blue and red sub-samples, we use the 
colour cut of \citet{Loveday2012}, namely
\begin{equation}
^{0.1}(g-r)_{\rm Kron} = 0.15 - 0.03\ ^{0.1}(M_r - 5 \log h).
\label{eqn:colourcut}
\end{equation}
A detailed investigation of colour bimodality in GAMA has recently 
been presented by \citet{Taylor2015}.
They utilise restframe and dust-corrected $(g - i)$ colour,
and argue that a probablistic assignment of galaxies to 'R' and 'B'
populations is preferable to a hard (and somewhat arbitrary) red/blue cut.
They also emphasise that colour is not synonymous with morphological type, 
but rather provides a proxy for mean stellar age within a galaxy.
Also, of course, a galaxy may appear red in uncorrected restframe colour
due to dust extinction, rather than an old stellar population.
In this paper, we stick with the simple colour-cut of 
equation~(\ref{eqn:colourcut}) for two reasons:
(i) to allow direct comparison with the results of \citet{Loveday2012};
(ii) the \citet{Taylor2015} model of the colour--mass distribution has
been tuned to a nearly volume-limited sample of galaxies at redshift
$z < 0.12$ --- the model parameters are likely to evolve at higher redshift.

Uncertainties in measured quantities, such as radial overdensity and the LF, 
are determined by jackknife resampling.
We subdivide the GAMA-II area into nine $4 \times 5$ degree regions,
and then recalculate the quantity nine times, omitting each region in turn.
For any quantity $x$, we may then determine its variance using
\begin{equation}
{\rm Var}(x) = \frac{N-1}{N} \sum_{i=1}^N (x_i - \bar{x})^2,
\end{equation}
where $N=9$ is the number of jackknife regions, $x_i$ is our estimate of $x$
obtained when omitting region $i$, and $\bar{x}$ is the mean of 
the $x_i$.
The numerator $(N-1)$ in the pre-factor allows for the fact that the
jackknife estimates are not independent.
Each jackknife region contains an average of 19,281 galaxies for the
full GAMA-II sample (i.e. without colour selection).

\section{Parametrizing the evolution} \label{sec:ev_model}

We parametrize luminosity and density evolution 
over the redshift range $0.002 < z < 0.65$
using the parameters 
$Q_e$ and $P_e$ introduced by \citet{Lin1999}.
This model assumes that galaxy populations evolve linearly with redshift
in absolute magnitude, parametrized by $Q_e$, and in log number density, 
parametrized by $P_e$.
Specifically, the luminosity $e$-correction is given by
$E(z) = Q_e(z - z_0)$, such that absolute magnitude $M$
is determined from apparent magnitude $m$ using
\begin{equation} \label{eqn:dmod}
  M = m - 5 \log_{10} d_L(z) - 25 - K(z; z_0) + Q_e(z - z_0),
\end{equation}
where $d_L(z)$ is the luminosity distance 
(assuming the cosmological parameters specified in the Introduction)
at redshift $z$ and $K(z; z_0)$ is the
$K$-correction, relative to a passband blueshifted by $z_0$.
Luminosity evolution is determined relative to the same redshift $z_0 = 0.1$
as the $K$-correction.

Evolution in number density $P(z)$ is parametrized as
\begin{equation} \label{eqn:den_ev}
  P(z) = P(z_0) 10^{0.4 P_e (z - z_0)} = P(z=0) 10^{0.4 P_e z}.
\end{equation}
The motivation for this choice of parametrization is that if the shape 
of the LF does not evolve with redshift, that is it shifts only horizontally
in absolute magnitude by $Q_e$, and vertically in log-density by $P_e$, 
then luminosity density $\rho_L$ evolves as 
\begin{equation} \label{eqn:lumden}
\rho_L(z) = \rho_L(z_0) 10^{0.4(P_e+Q_e)(z-z_0)}.
\end{equation}
While $P_e$ and $Q_e$ are strongly degenerate, and so poorly constrained
individually, their sum $P_e+Q_e$ is well-constrained
\citep{Lin1999,Loveday2012}.
We set further constraints on the linear combination of these parameters
in Section~\ref{sec:results}.

\section{Density-corrected \Vmax\ method} \label{sec:Vmax}

In this section we describe our technique for determining the LF using
a maximum-likelihood, density-corrected \Vmax\ estimator, assuming
that evolution is known.
We will discuss how we determine the evolution parameters $Q_e$ and $P_e$
in Section~\ref{sec:ev}.
Our method is based on the joint stepwise maximum likelihood
(JSWML) method of \citet{Cole2011}, which jointly fits the LF and
overdensities in radial bins of redshift caused by large-scale structure.
Cole's derivation starts with an expression for the {\em joint} probability
of finding a galaxy at specified redshift and luminosity,
and assumes that all galaxies have identical evolution- and $K$-corrections.
We wish to allow for individual $K$- (and in the future $e$-)
corrections, in which case
it is easier to start with the {\em conditional} probability that an observed
galaxy of luminosity $L_i$ has a redshift $z_i$, 
assuming that the luminosity and spatial dependence 
of the galaxy number density are separable.
This conditional probability is given by \citep{Saunders1990}:
\begin{equation}
  \label{eqn:p}
  p_i = \frac{\Delta(z_i) P(z_i) \left. \frac{dV}{dz}\right|_{z_i}}
  {\int_0^{z_{{\rm max}, i}} \Delta(z) P(z) \frac{dV}{dz} dz}.
\end{equation}
Here we have factored the mean density at redshift $z$,
$\bar{n}(z) = \Delta(z) P(z)$, into a product of
the galaxy overdensity\footnote{Following \citet{Cole2011},
we use the term overdensity to mean a multiplicative relative density,
so that $\Delta = 1$ corresponds to average density.} 
$\Delta(z)$ due to large-scale structure
times the steadily evolving density $P(z)$ from equation~(\ref{eqn:den_ev});
$dV/dz$ is the differential of the survey volume, and
$z_{{\rm max}, i}$ is the maximum redshift at which galaxy $i$
would still be visible, determined by the survey flux limit 
along with the galaxy's luminosity, $K$- and $e$-corrections.

Adopting binned estimates of the galaxy overdensity $\Delta$,
and weighting each galaxy by its incompleteness-correction weight, $W_i$,
we obtain a log-likelihood
\begin{equation}
  \label{eqn:log-like}
  \ln {\cal L} = \sum_i W_i 
  \left[ \ln \sum_j \Delta_j P_j V_j D_{ij}
   - \ln \sum_j \Delta_j P_j V_j S_{ij} \right].
\end{equation}
Here $V_j$, $P_j$ and $\Delta_j$ are the volume, density evolution
and galaxy overdensity respectively in redshift bin $j$; 
the function $D_{ij}$ is a simple binning function, equal to unity
if galaxy $i$ lies in redshift bin $j$, zero otherwise,
and $S_{ij}$ is the fraction of redshift bin $j$ in which galaxy $i$ is visible.
In the present analysis we employ redshift bins of width $\Delta z = 0.01$.
The maximum-likelihood solution for the overdensities $\Delta_j$, given by
$\partial \ln {\cal L}/\partial \Delta_j = 0$, 
may be obtained by iteration from:
\begin{equation}
  \label{eqn:delta}
  \Delta_j = W_{{\rm sum},j} \left[\sum_i 
  \frac{W_i P_j V_j S_{ij}}{\Vdcmax_i}
  \right]^{-1},
\end{equation}
where $W_{{\rm sum},j} = \sum_i W_i D_{ij}$ is the sum of galaxy weights 
in redshift bin $j$ and $\Vdcmax_i = \sum_k \Delta_k P_k V_k S_{ik}$,
the effective volume, 
corrected for evolution and fluctuations in radial density, 
within which galaxy $i$ is visible.

The LF, unaffected by density fluctuations, may then be estimated by 
substituting \Vdcmax for the usual expression for \Vmax:
\begin{equation}
  \label{eq:phi}
  \phi^{\rm bin}_l = \sum_i \frac{W_i D_{il}}{\Vdcmax_i},
\end{equation}
where $D_{il} = 1$ if galaxy $i$ is in luminosity bin $l$, zero otherwise.
\citet{Cole2011} shows that this expression may be derived via maximum
likelihood, at least in the case of identical $e$- and $K$-corrections.

Cole also discusses an extension to this method whereby parameter(s) describing
the density evolution $P(z)$ may be determined simultaneously with the
overdensities $\Delta_j$ by adding prior constraints on the values of
$\Delta_j$ using the known clustering of galaxies.
However, for our choice of density evolution parametrization
(equation~\ref{eqn:den_ev}),
the derivative in Cole equation~(25) no longer depends explicitly 
on the evolution parameter, leading to a lack of convergence.
We therefore prefer to search over both luminosity and density evolution 
parameters, as described in the next section.

A stepwise estimate of the LF, as given by equation~(\ref{eq:phi}),
is not constrained to vary smoothly from bin to bin.  
Furthermore, at very low and high luminosity there may be bins containing 
no galaxies, resulting in an ill-defined log-likelihood 
(see equation~\ref{eqn:post} below).
This problem is exacerbated when exploring possible values of the
luminosity evolution parameter $Q_e$, 
as galaxies will then shift from bin to bin
as $Q_e$ is varied, resulting in unphysical sharp jumps in likelihood.
To overcome these problems, we employ a Gaussian-smoothed estimate of the LF:
\begin{equation}
  \label{eq:phi_kde}
  \phi^{\rm GS}_l =  \sum_i \frac{W_i}{\Vdcmax_i} G \left(\frac{M_i - M_l}{b} \right).
\end{equation}
Here the smoothing kernel $G$ is a standard Gaussian, 
$b$ is the smoothing bandwidth,
$M_i$ is the ($K$- and $e$-corrected) absolute magnitude of galaxy $i$ and
$M_l$ is the absolute magnitude at the centre of bin $l$.
In order not to underestimate the extreme faint-end of the LF,
it is important to apply boundary conditions to $\phi^{\rm GS}$ corresponding 
to the chosen range of absolute magnitudes.
We do this using the default renormalization method and bandwidth choice
of the python module 
\verb|pyqt_fit.kde|\footnote{\url{https://pypi.python.org/pypi/PyQt-Fit}}.
$\phi^{\rm GS}$ does not, of course, correspond to the true galaxy LF, 
but rather to the LF convolved with a Gaussian of standard deviation $b$.
Therefore when plotting the LF and fitting a Schechter function, 
we use the standard binned LF $\phi^{\rm bin}$ rather than $\phi^{\rm GS}$.

\section{Determining evolution parameters} \label{sec:ev}

In Cole's original derivation of this method, one maximises a
posterior likelihood (Cole equation 38)\footnote{Note that Cole equations
(36--38) are missing factors of $P_p$, such that each occurrence of $V_p$
should read $V_p P_p$.}
over the luminosity evolution parameter $Q_e$ (Cole calls this parameter $u$).
When applying this method to GAMA data, we found that the estimated
value of $Q_e$ diverged, unless one places an extremely tight prior on
its value\footnote{
We believe that the reason that the test described in Section 5 of 
\citet{Cole2011} was successful was due to (i) placing a very tight prior
($\sigma = 0.05$) on the density evolution parameter, and (ii) simulating a very
deep galaxy survey (extending to magnitude $r < 24$ and redshift $z < 1.4$).
Both of these factors minimize the degeneracy between luminosity and
density evolution, and hence aid convergence.
The GAMA-II sample is significantly shallower ($r < 19.8$, $z < 0.65$),
and we do not wish to place tight prior constraints on either of the evolution
parameters.
}.
Our problem was traced to the fact that varying $Q_e$ changes all of the
inferred absolute magnitudes (as well as visibility limits) for each galaxy.
Choosing fixed absolute magnitude limits within which to determine the LF
thus results in a change of sample size as $Q_e$ varies, leading to 
likelihoods that cannot be directly compared.
Even if one includes the term on the second line of Cole equation~(36),
which yields $-N_{\rm tot} \ln \hat{N}_{\rm tot}$ in the case of identical $K$-
and $e$-corrections, the estimate of $Q_e$ 
still diverges as galaxies
shift systematically brighter or fainter as $Q_e$ decreases or increases.
We therefore consider two alternative methods to optimize the
evolution parameters.

\subsection{Mean probability}

Our first solution is to consider not the product of the probabilities of
observing each galaxy, but instead the {\em geometric mean} 
of the probabilities which does not vary systematically with
sample size $N$.
Our pseudo-log-likelihood $\ln {\cal P}$ is then given by (Cole equation 36)
\begin{equation} \label{eqn:post}
  \begin{split}
  \ln {\cal P} &= \frac{1}{N} \sum_j W_{{\rm sum},j} \ln(V_j P_j \Delta_j)
  + \frac{1}{N} \sum_l W_{{\rm sum},l} \ln \phi^{\rm GS}_l \\
  &- \frac{1}{N}\sum_i W_i \ln \sum_j V_j P_j \Delta_j \sum_l \phi^{\rm GS}_l 
  S(L_{{\rm min},i, j} | L_l) \\
  &- \sum_j \frac{(\Delta_j -1)^2}{2 \sigma^2_{\Delta_j}}
  - \frac{(Q_e - Q_0)^2}{2 \sigma^2_{Q_e}}
  - \frac{(P_e - P_0)^2}{2 \sigma^2_{P_e}},
  \end{split}
\end{equation}
where $W_{{\rm sum},j}$ is the sum of galaxy weights in redshift bin $j$,
$W_{{\rm sum},l}$ is the sum of galaxy weights in luminosity bin $l$,
and $S(L_{{\rm min},i, j} | L_l)$ is the fraction of luminosity bin $l$
for which galaxy $i$ at redshift $z_j$ would be visible.
The term on the second line is a constant in the case of identical
$K$- and $e$-corrections; with identical $e$- but independent $K$-corrections
we find that including this term makes a negligible difference to the
maximum-likelihood solution.
The terms on the third line are priors on the radial overdensities $\Delta_j$
and the evolution parameters $P_e$ and $Q_e$.
The priors on $\Delta_j$ are essential, as these values are 
completely degenerate with the density evolution parameter $P_e$.
As discussed by Cole, the expected variance in $\Delta_j$ is given by
\begin{equation} \label{eqn:den_var}
  \langle \sigma^2_{\Delta_j} \rangle = \frac{1 + 4 \pi \hat{n}_j J_3}{\hat{n}_j V_j},
\end{equation}
with $\hat{n}_j$ the predicted density and $V_j$ the volume of redshift bin $j$.
The factor $J_3 = \int r^2 \xi(r) dr \approx 2,000 h^{-3} {\rm Mpc}^3$
accounts for the fact that because galaxies 
are clustered, they tend to come in clumps of $4\pi \hat{n} J_3$ 
galaxies at a time
\citep{Peebles1980}.
We find, however, that much more reliable estimates of $\sigma^2_{\Delta_j}$
are obtained from jackknife sampling --- see Fig.~\ref{fig:delta} below.
This is particularly true in the higher redshift bins, where one is sampling
the clustering of the most luminous galaxies, and where adopting a universal
value for $J_3$ underestimates the actual density fluctuations observed
between jackknife samples.
The priors on $P_e$ and $Q_e$ are optional, 
and may help convergence in some cases.
We adopt broad priors of $(Q_0, \sigma^2_{Q_e}) = (1, 1)$ and
$(P_0, \sigma^2_{P_e}) = (2, 1)$.
These values were chosen to be consistent with the findings of 
\citet{Loveday2012} while still allowing some freedom for the optimum values
to change under the present analysis.

\subsection{LF--redshift $\chi^2$}

Our second method compares LFs estimated in two or more redshift ranges:
if the evolution and density variations are correctly modelled, then the LFs
should be in good agreement; if evolution parameters are poorly estimated, 
then one would expect poor agreement.
We then minimize the $\chi^2$ 
($\equiv -2 \ln {\cal L}$)
given by
\begin{equation} \label{eqn:lf_chi}
  \chi^2 = \sum_{j, k > j} \sum_l \frac{(\phi^j_l - \phi^k_l)^2}
  {\Var(\phi^j_l) + \Var(\phi^k_l)} + 
  \sum_j \frac{(\Delta_j -1)^2}{\sigma^2_{\Delta_j}},
\end{equation}
where $\phi^j_l$ is the 
Gaussian-smoothed
LF in magnitude bin $l$ for the broad redshift range $j$,
and $\Var(\phi^j_l)$ is the corresponding variance, determined by jackknife 
resampling.
We restrict the sum over magnitude bins $l$ to those bins which are complete
given the redshift limits (see Section 3.3 of \citealt{Loveday2012}) 
and which include at least ten galaxies
for all values of $Q_e$ between specified limits.
In practice, we have found best results are achieved using just two redshift
ranges, split near the median redshift of the sample, $\bar{z} \approx 0.2$,
so that the `knee' region of the LF around $L^*$ is well-sampled by both,
and hence the degeneracies between luminosity and density evolution are 
minimized.
If one chooses three or more redshift ranges, there will be very little
luminosity coverage in common to the lowest and highest ranges,
and so one does not really gain much information in doing so.
Again, it is essential to place a prior on the overdensities
(final sum in equation~\ref{eqn:lf_chi}, with $\sigma^2_{\Delta_j}$ 
also determined from jackknife resampling) to remove the degeneracy 
with density evolution.
This method places no priors on the values for the evolution parameters.

\subsection{Finding optimum evolution parameters}

We first evaluate $\chi^2$ values, using each of the above methods,
on a rectangular grid of $(P_e, Q_e)$, thus allowing one to visualise 
the correlations between the evolution parameters.
The grid point with the smallest $\chi^2$ value is then used as 
a starting point for a downhill simplex minimisation to refine the 
parameter values corresponding to minimum $\chi^2$.

In order to quantify the degeneracy between evolution parameters,
we slice the $\chi^2$ grid in bins of $P_e$.
For each slice we fit a quadratic function to $\chi^2(Q_e)$ 
using the five $(Q_e, \chi^2)$ values closest to the point of minimum $\chi^2$
in that slice.
Using this quadratic fit, we locate the point $Q_{e, \chi^2_{\rm min}}$
of minimum $\chi^2$ and its 1-sigma range,
i.e. the range of $Q_e$ values where $\chi^2$ increases by unity 
from the minimum.
We find both for simulations and for real data that the 
$Q_{e, \chi^2_{\rm min}}$--$P_e$
relation is very well fit by a straight line, and so
we perform a linear least-squares fit to $(P_e, Q_{e, \chi^2_{\rm min}})$
to obtain the relation $Q_e = mP_e + c$ which minimizes $\chi^2$.

\section{Tests using simulated data} \label{sec:tests}

\subsection{The simulations}

In this section, we test our implementation of the JSWML estimator using 
simulated data, following the procedure outlined in Section~5 of 
\citet{Cole2011}.

We start by choosing a model LF with \citet{Schechter1976} 
and evolution parameters 
close to those obtained from the GAMA-I survey by \citet{Loveday2012} 
and as given in Table~\ref{tab:sims}.
We then randomly generate redshifts with a uniform density in comoving
coordinates, modulated by our assumed density evolution 
(equation~\ref{eqn:den_ev}), over the range $0.002 < z < 0.65$.
Absolute magnitudes are selected randomly according to our assumed Schechter
function from the range $-24 < M < -12$.
From each absolute magnitude we subtract $Q_e (z-0.1)$ 
to model luminosity evolution.
We then assign apparent magnitudes $r$ using $K$-correction coefficients 
selected randomly from the GAMA-II data and reject simulated galaxies 
fainter than $r = 19.8$.
This process is repeated until sufficient random galaxies have been generated 
to give the required number density,
\begin{equation}
N_{\rm sim} = \int_{z_{\rm min}}^{z_{\rm max}} 
\int_{L_{\rm min}(z, m_{\rm min})}^{L_{\rm max}(z, m_{\rm max})}
\phi(L,z) dL \frac{dV}{dz} dz,
\end{equation}
within a volume corresponding to that of the three GAMA-II fields, viz 
$3 \times 5^\circ \times 12^\circ = 180$ deg$^2$.

In order to simulate the effects of galaxy clustering, we spilt the
simulated volume into 65 redshift shells $p$ of equal thickness 
$\Delta z \approx 0.01$ and with volume $V_j$.
For each shell we generate a random density perturbation $\delta_j$
drawn from a Gaussian with zero mean and variance $4 \pi J_3 / V_j$,
with $4 \pi J_3 = 30,000 \volunit$.
We then randomly resample $N'_j = (1 + \delta_j) N_j$ of the original $N_j$
simulated galaxies in each shell $p$, thus producing fluctuations
consistent with the assumed value of $J_3$.

Imaging completeness and redshift success are modelled by
generating surface brightnesses and fibre magnitude for each
simulated galaxy according to the relations observed in GAMA-I data,
see Appendix~A1 of \citet{Loveday2012}.
Imaging completeness $C_{\rm im}$ is then determined from Fig.~1 of 
\citet{Loveday2012}
and redshift success $C_z$ from equation~(\ref{eq:zcomp}).
Simulated galaxies are then chosen randomly with probability equal to
$C_{\rm im} C_z$ and assigned a weight
$W_i = 1/({C_{\rm im}}_i {C_z}_i)$ to compensate for those
simulated galaxies omitted from the sample.

This procedure is repeated to generate ten independent 
mock catalogues, each containing around 180,000 galaxies.
These mock catalogues are run through the JSWML estimator, with
evolution parameters being determined using both methods discussed
in the previous section.
Since the mock galaxies are clustered only in redshift shells and not in
projected coordinates on the sky, we determine
the expected variance in overdensity using equation~(\ref{eqn:den_var})
rather than jackknife resampling.
We employ 65 redshift shells out to $z = 0.65$ and calculate the LF
in bins of $\Delta M = 0.25$ mag over the range $-23 < M < -15$ mag.

\subsection{Simulation results}

\begin{table}
 \caption{Mean and standard deviation of the evolution and Schechter 
parameters recovered from ten simulated GAMA catalogues.
Parameters $m$ and $c$ quantify the linear relation $Q_e = mP_e + c$ 
which minimizes $\chi^2$.
}
 \label{tab:sims}
 \begin{math}
 \begin{array}{rrrr}
 \hline
 & \mbox{True} & \multicolumn{1}{c}{\mbox{mean prob}} & 
\multicolumn{1}{c}{\mbox{LF--redshift }\chi^2} \\
 \hline
 Q_e & 0.7 & 0.68 \pm 0.23 & 0.61 \pm 0.14 \\
 P_e & 1.8 & 1.71 \pm 0.67 & 1.88 \pm 0.54 \\
 \Cov(Q_e,P_e) & & -0.16 & -0.08 \\
 m &  & -0.37 \pm 0.01 & -0.33 \pm 0.01 \\
 c &  & 1.31 \pm 0.09 & 1.24 \pm 0.07 \\
 \alpha & -1.23 & -1.23 \pm 0.02 & -1.23 \pm 0.01 \\
 M^* - 5 \lg h & -20.70 & -20.72 \pm 0.06 & -20.74 \pm 0.03 \\
 \lg (\phi^* / \denunit) & -2.00 & -2.10 \pm 0.09 & -2.12 \pm 0.07 \\
 \hline
 \end{array}
 \end{math}
\end{table}

\begin{figure}
\includegraphics[width=\linewidth]{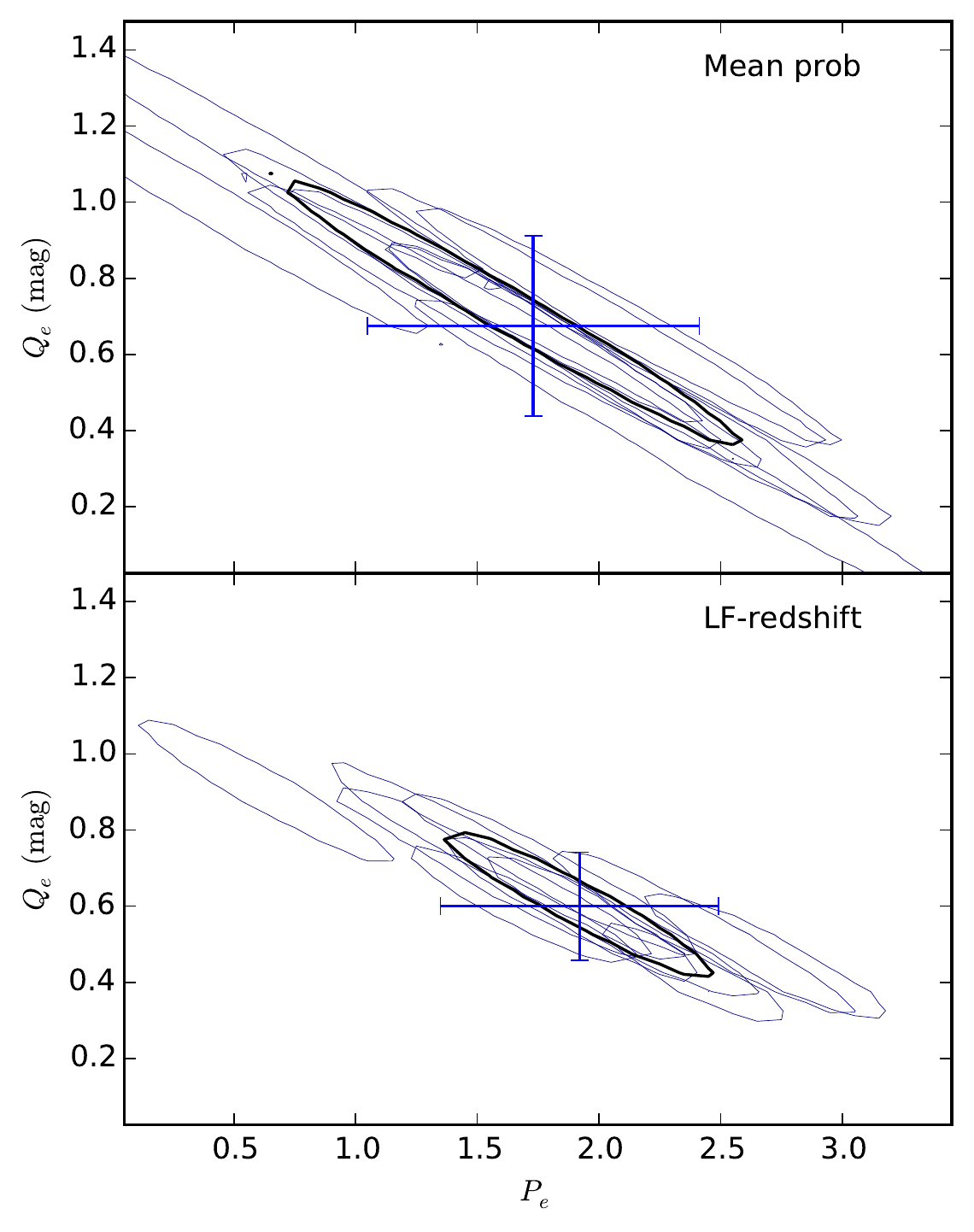}
\caption{95 per cent confidence limits on evolution parameters determined from
ten simulated datasets (light contours) and their average (heavy contour)
determined using (top) mean probability
(equation~\ref{eqn:post}) and (bottom) LF--redshift $\chi^2$ 
(equation~\ref{eqn:lf_chi}).
The error bars show the mean and standard deviation of the $(P_e, Q_e)$ 
parameters
from each simulation which yield  minimum $\chi^2$.
The input evolution parameters for these simulations were 
$P_e = 1.8$, $Q_e = 0.7$.
}
\label{fig:like_sim}
\end{figure}

The mean and standard deviation of each recovered parameter,
and the covariance between evolution parameters,
are given in Table~\ref{tab:sims}.
We see that the input evolution and LF parameters are recovered within about 
one standard deviation for both methods.

Fig.~\ref{fig:like_sim} shows 95 per cent confidence limits on the 
evolution parameters measured from each of the simulations.
We see that the error contours are significantly smaller using the 
LF--redshift $\chi^2$ method compared with the mean probability method.
However, this test is idealized, in that our choice of evolution 
parametrization 
is identical in the simulations and in the analysis\footnote{
We are performing a self-consistency test.
It is unlikely that real galaxy populations evolve exactly according
to our parametrization.
},
and so we will apply both
methods of constraining evolution parameters to the GAMA data in the 
following Section.
Note that the simulations have no inbuilt covariance between evolution 
parameters: they all use identical values of $P_e$ and $Q_e$.
The degeneracies (as quantified by $\Cov(Q_e,P_e)$ and the parameters 
$m$ and $c$ 
in Table~\ref{tab:sims}) arise as a result of the fitting process.
For an LF described by an unbroken power law, the degeneracy between 
$P_e$ and $Q_e$
would be total, i.e. evolution in luminosity and density would be
indistinguishable.

\section{Results from GAMA} \label{sec:results}

\subsection{Evolution}

\begin{table*}
\caption{Best-fitting evolution parameters for GAMA-II 
galaxy samples obtained using both mean probability and LF--redshift methods.
Parameters $m$ and $c$ quantify the linear relation $Q_e = mP_e + c$.
For the LF--redshift method only, $\chi^2_\nu$
is the reduced $\chi^2$ from equation~(\ref{eqn:lf_chi});
The uncertainties quoted on $Q_e$ and $P_e$ come from the bounding box
containing the 1-sigma likelihood contour; the uncertainty on $P_e+Q_e$ is given
by the distance from the point of minimum $\chi^2$ to the 1-sigma 
likelihood contour along the direction $P_e = Q_e$.
}
\label{tab:ev_fits}
\input ev_table.tex
\end{table*}

\begin{figure}
\includegraphics[width=\linewidth]{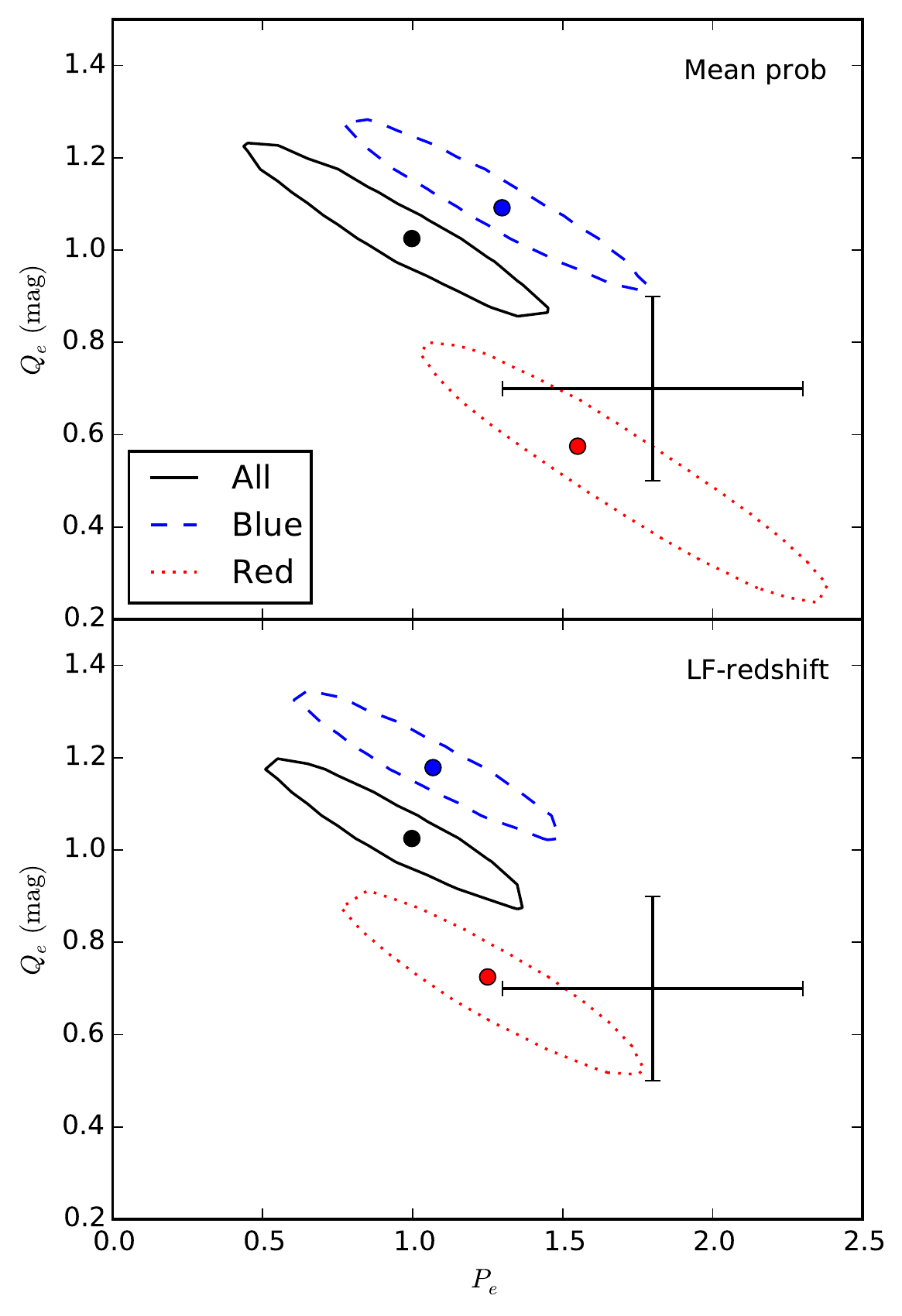}
\caption{95 per cent confidence limits on GAMA-II evolution 
parameters for all, blue and red galaxies as labeled.
The upper panel shows the limits obtained using mean probability
(equation~\ref{eqn:post}); the lower panel shows results
using LF--redshift $\chi^2$ (equation~\ref{eqn:lf_chi}).
The large dots indicate the location of minimum $\chi^2$.
The large errorbars show the evolution parameters and 68 per cent 
confidence limits estimated for the
combined GAMA-I sample in the $r$ band by 
\citet[$Q_{\rm par}$ and $P_{\rm par}$ from Table~5]{Loveday2012}.
}
\label{fig:chigrid}
\end{figure}

Fig.~\ref{fig:chigrid} shows 95 per cent confidence limits on the 
evolution parameters $P_e, Q_e$ determined using equations~(\ref{eqn:post})
and (\ref{eqn:lf_chi}) 
for the full GAMA-II sample and for blue and red galaxies separately.
We see that the confidence limits obtained with the two different methods
largely overlap, although there are small differences between them.
Best fit evolution parameters are given in Table~\ref{tab:ev_fits}.
The difference in LFs obtained using evolution parameters determined
with the two different methods is negligible 
(much less than the 1-$\sigma$ random errors;
see Table~\ref{tab:lf_fits}).
This illustrates the robustness of the LF estimate to the individual
values assumed for $P_e$ and $Q_e$: 
as long as their joint estimate is reasonable,
e.g. they lie within the 95 per cent likelihood contours of 
Fig.~\ref{fig:chigrid},
then overestimating one evolution parameter (e.g. $P_e$) 
is largely compensated for by underestimating the other (e.g. $Q_e$).

The differences in density evolution ($P_e$) for
red and blue galaxies are not significant.
Blue galaxies do however exhibit significantly stronger evolution 
in luminosity ($Q_e$) and in luminosity density ($Q_e + P_e$) than red galaxies,
at the $\sim 5$-$\sigma$ level.

The differences between red and blue galaxies agree qualitatively with those 
of \citet{Loveday2012}, although in the present analysis we no longer
see any evidence for negative density evolution for red galaxies.
The three samples show very similar degeneracies in $(P_e, Q_e)$ 
parameter space.
The errors on $P_e$ and $Q_e$ in Table~\ref{tab:ev_fits} are the formal errors
obtained by holding one parameter fixed and varying the other until $\chi^2$ 
increases by one.
Given the scatter in 95 per cent confidence limits between simulations
shown in Fig.~\ref{fig:like_sim}, more realistic errors, and their covariance,
may be obtained from Table~\ref{tab:sims}.

Since the exact values assumed for the evolution parameters 
have such a small effect on the LF parameters, 
see Table~\ref{tab:lf_fits} below,
for the remainder of this paper we assume evolution parameters found from the
LF--redshift method in the lower half of Table~\ref{tab:ev_fits}.

\subsection{Radial overdensities}

\begin{figure}
\includegraphics[width=\linewidth]{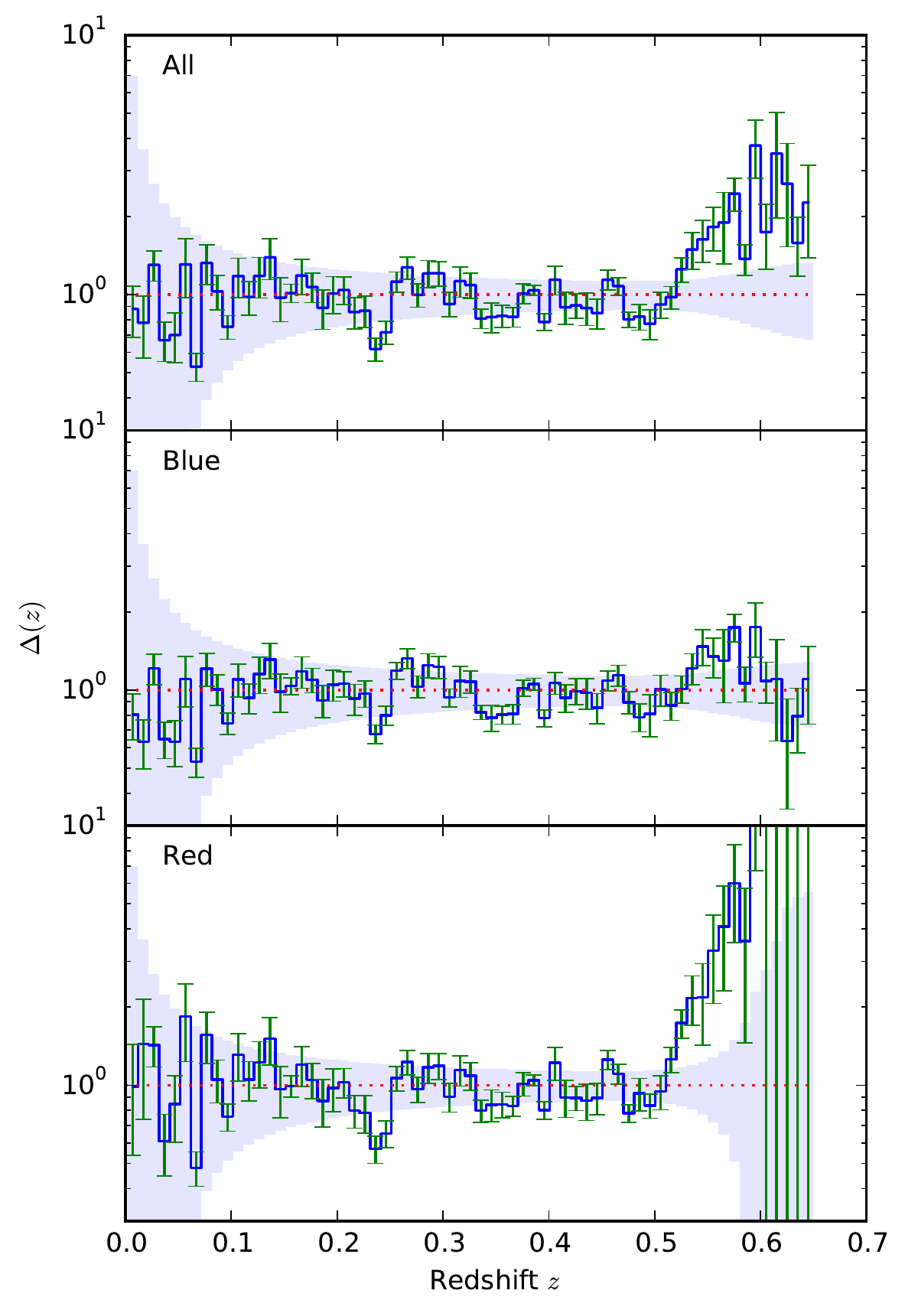}
\caption{Radial overdensities determined from GAMA-II using the entire sample
and blue and red subsets as labelled, assuming evolution parameters 
as given in the lower half of Table~\ref{tab:ev_fits}.
The error bars show uncertainties estimated from jackknife sampling 
and the shaded regions centred on $\Delta = 1$ show the expected variance from 
equation~(\ref{eqn:den_var}).
}
\label{fig:delta}
\end{figure}

\begin{figure}
\includegraphics[width=\linewidth]{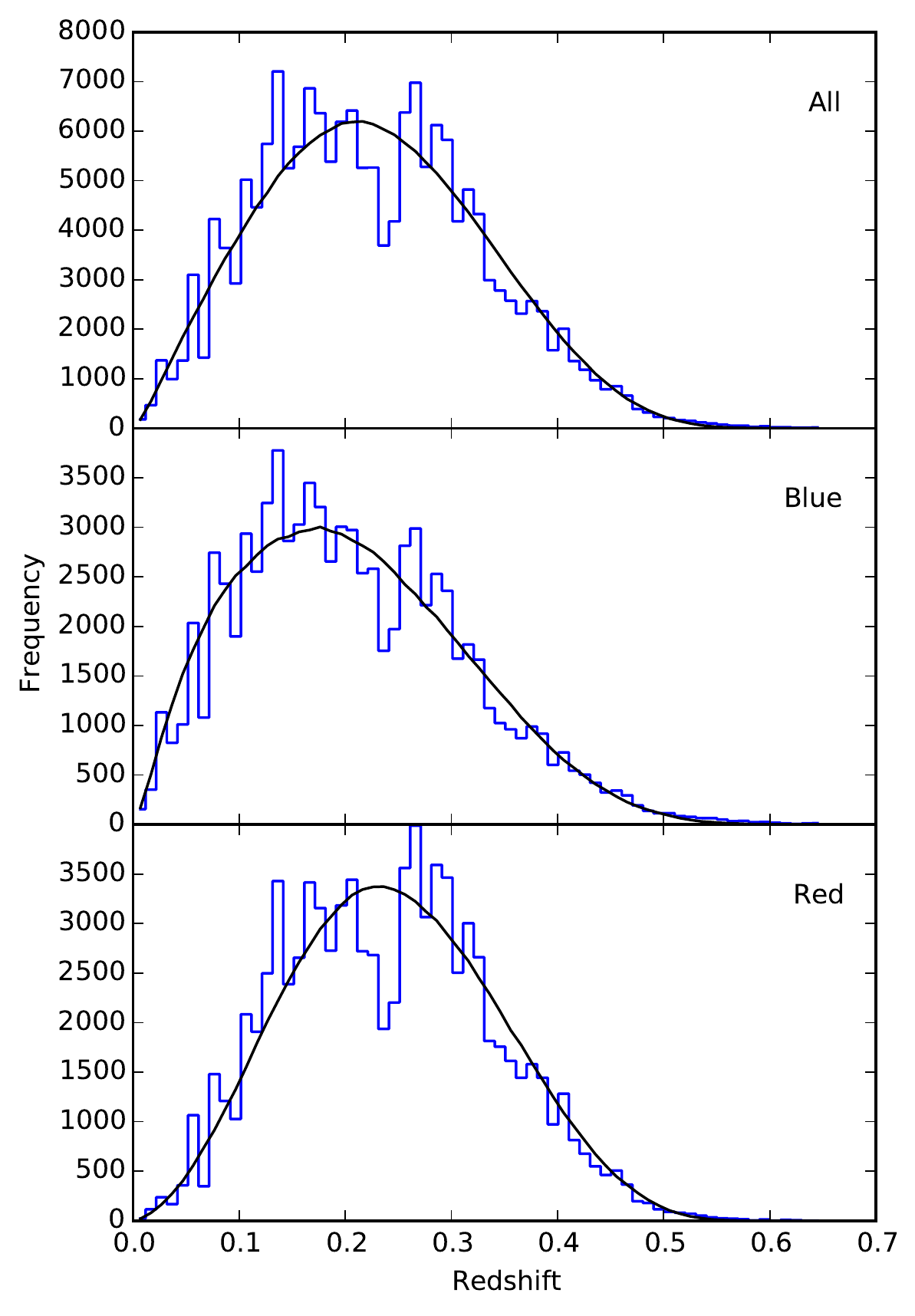}
\caption{Redshift histograms for the whole GAMA-II sample and for 
blue and red galaxies separately.
The curves in each panel give the predicted redshift distribution
based on our evolving LF model fits.
}
\label{fig:Nz}
\end{figure}

Radial overdensities are shown in Fig.~\ref{fig:delta}.
While our evolution model is performing well, insofar as $\Delta(z)$ oscillates 
about unity, for redshifts $z \la 0.5$, beyond this limit the 
overdensities are systematically high.
This effect is almost entirely due to red galaxies, 
suggesting that luminosity and/or density evolution increases sharply
at $z \approx 0.5$ for these galaxies compared with our model 
(section~\ref{sec:ev_model}).
It seems unlikely that incompleteness corrections could cause this, 
as there is no noticeable increase in weights beyond $z = 0.5$.
Only 0.8 per cent of GAMA-II main survey galaxies lie beyond $z = 0.5$,
too few to constrain a more complicated evolution model, or to look
for a large overdensity at these redshifts,
see Fig.~\ref{fig:Nz}.

Below redshifts $z = 0.5$, 
we see the same features in radial overdensity in all three samples,
although the fluctuations, as expected, are slightly more pronounced
in the red galaxy sample.
Note that the error bands given by equation~(\ref{eqn:den_var}) 
(shaded regions in Fig.~\ref{fig:delta}) are 
significantly larger/smaller than the jackknife errors at low/high redshift.
There are two reasons for this:
(i) the low-redshift bins sample too small a volume
for the $J_3$ integral to have converged, and (ii) the low/high-redshift 
bins are dominated by faint/luminous galaxies, with weaker/stronger clustering
than the average defined by the assumed value of $J_3$.
This is why we use jackknife errors rather than the predicted variance
in determining $\sigma^2_{\delta_j}$.
We have tried halving the number of redshift bins to 32,
verifying that the fitted parameters are insensitive to the
redshift binning, with parameters changing by less than one sigma 
when the redshift bin size is doubled from $\Delta z = 0.01$ to 
$\Delta z = 0.02$.

\subsection{LFs}

\begin{figure}
\includegraphics[width=\linewidth]{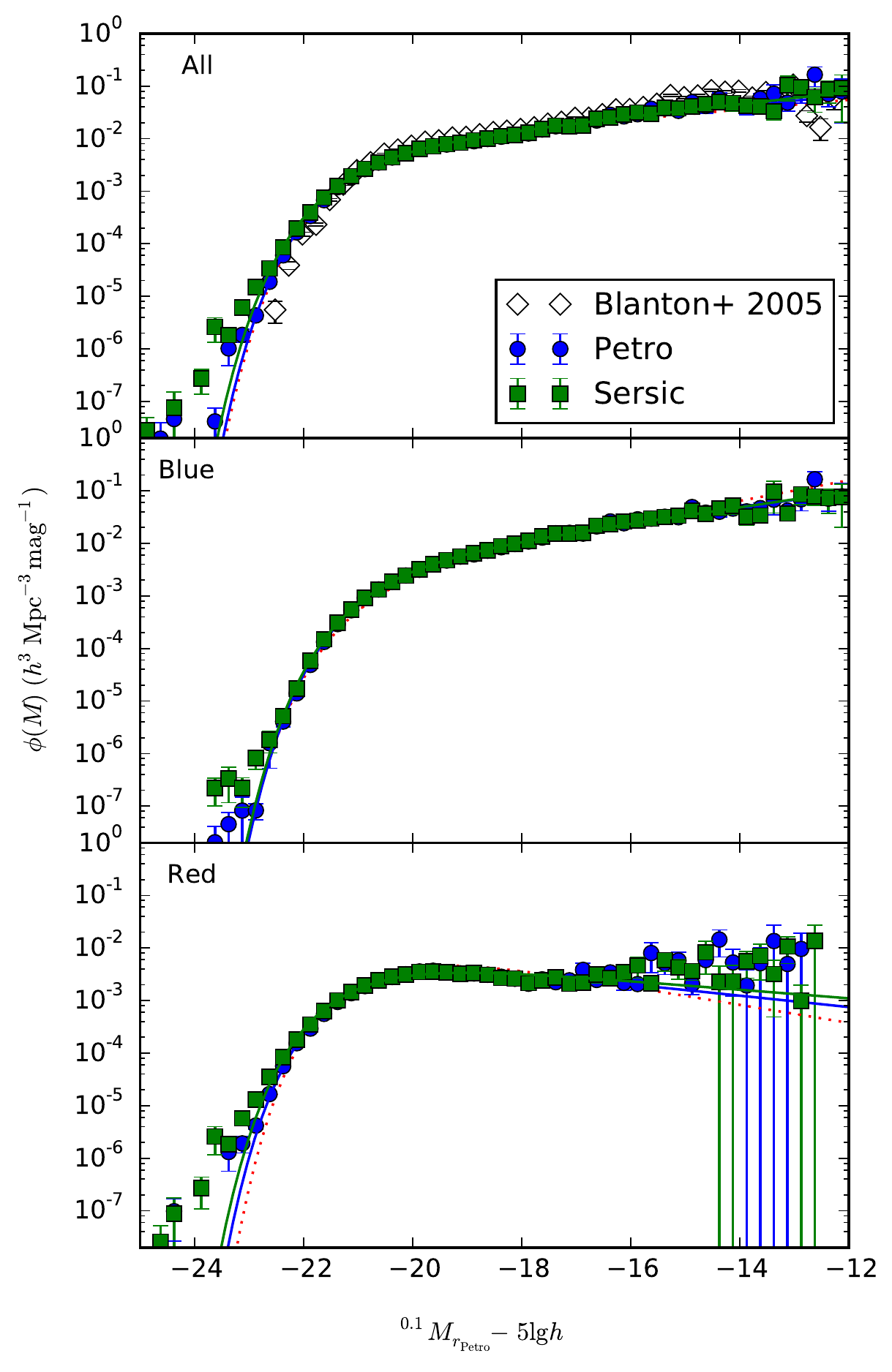}
\caption{GAMA-II evolution- and density-corrected Petrosian (blue circles)
and \Sersic\ (green squares) $r$-band LFs with 
best-fitting Schechter functions (solid lines) assuming evolution parameters 
for each sample as given in the lower half of Table~\ref{tab:ev_fits}.
The dotted lines show the best-fit $r$-band Schechter functions from Table~5
of \citet{Loveday2012}.
The open diamonds in the top panel show the `corrected' LF from Fig.~7 of
\citet{Blanton2005}.
}
\label{fig:lfr}
\end{figure}

\begin{table}
\caption{Best-fitting $r$-band LF parameters for GAMA-II 
galaxy samples obtained using both mean probability and LF--redshift methods.
For the latter method, we show LF parameters obtained using both Petrosian and
\Sersic\ magnitudes.
$\chi^2_\nu$ is the reduced $\chi^2$ from least-squares 
Schechter function fits to the LF estimates; 
none of the LFs are well-fit in detail by a Schechter function,
particularly at the bright end.
The uncertainties quoted on the LF parameters come from jackknife sampling,
but do not explicitly include the large degeneracies between them.
}
\label{tab:lf_fits}
\input lf_table.tex
\end{table}

Petrosian and \Sersic\ $r$-band LFs are shown in Fig.~\ref{fig:lfr}.
Surface brightness and redshift incompleteness have been taken into account by
appropriately weighting each galaxy 
(up to a maximum weight of 5.0, Section~\ref{sec:data}).
We have fit a Schechter function to each binned LF using least squares;
the fit parameters are tabulated in Table~\ref{tab:lf_fits}.
Note that the Schechter fit for red galaxies underestimates
the faint end of the LF
(as well as the bright end --- see below).
It is likely that the faint-end upturn for red galaxies
is at least partly due
to the inclusion of dusty spirals in this sample;
the luminosity and stellar mass functions of E--Sa galaxies of 
\citet{Kelvin2014,Kelvin2014a} show no indication of a faint-end or
low-mass upturn.
Fig.~5 of \citet{Kelvin2014a} shows that while very few galaxies with elliptical
morphology are blue ($(g-i)_0 \la 0.6$), 
the converse is not true: a substantial number of 
galaxies with spiral morphology are red ($(g-i)_0 \ga 0.8$).
Any upturn in the luminosity or mass function of spheroidal galaxies
is more likely to be due to the presence of so-called little blue spheroids
\citep[Fig.~A1]{Kelvin2014a}.
Finally, we note that
\citet{Taylor2015} have shown that the shape of the low-mass end of the
stellar mass function of red galaxies is sensitive to how 'red' is defined.
A low-mass upturn is seen when using the definition of \citet{Peng2010},
but not when using those of \citet{Bell2003} and \citet{Baldry2004}.

The red galaxy LF, and that for the combined sample, show a bright-end excess:
there are significantly more high-luminosity ($M_r - 5 \lg h < -23$ mag)
galaxies than predicted by the Schechter function fit.
This is particularly true for the LF measured using \Sersic\ magnitudes, 
which capture a larger fraction of the total light for de Vaucouleurs profile 
galaxies which dominate the bright end of the LF (e.g. \citealt{Bernardi2013}).
A bright-end excess above a best-fitting Schechter function has been observed
in many other surveys \citep[e.g.][]{Loveday1992,Norberg2002,Montero-Dorta2009}
and appears to be particularly pronounced in bluer bands
\citep[e.g.][]{Montero-Dorta2009,Loveday2012,Driver2013}.
As \citet{Driver2013} point out, given the approximately Gaussian distribution
of galaxy colours, the LF cannot be well fit by a Schechter function 
in all bands.
One should however be aware of the possibility that \Sersic\ magnitudes,
extrapolated as they are out to ten effective radii, are susceptible to over-
(or under-) estimating the flux of even isolated galaxies if the \Sersic\
parameters are poorly fit (although the fitting pipeline does attempt
to trap for poor fits).
Hence we also show LFs using more stable Petrosian magnitudes.

Our Schechter fits to these LFs are consistent with the $r$-band LFs 
determined from the GAMA-I sample by \citet{Loveday2012}, 
using slightly different methods, and shown in Fig.~\ref{fig:lfr} 
as dotted lines.
We also show the `corrected' LF from the \citet{Blanton2005} low-redshift
SDSS sample (without colour selection).
Considering that this plot is comparing the LFs of SDSS galaxies within 
only $150 \hMpc$ with GAMA galaxies out to $z \approx 0.65$, the agreement 
is remarkably good, and provides further evidence that 
the simple evolutionary model adopted allows one to accurately recover 
the evolution-corrected LF,
despite its poor performance beyond redshift $z \approx 0.5$
(Fig.~\ref{fig:delta}).

\subsection{Testing the evolution model} \label{sec:ev_test}

\begin{figure}
\includegraphics[width=\linewidth]{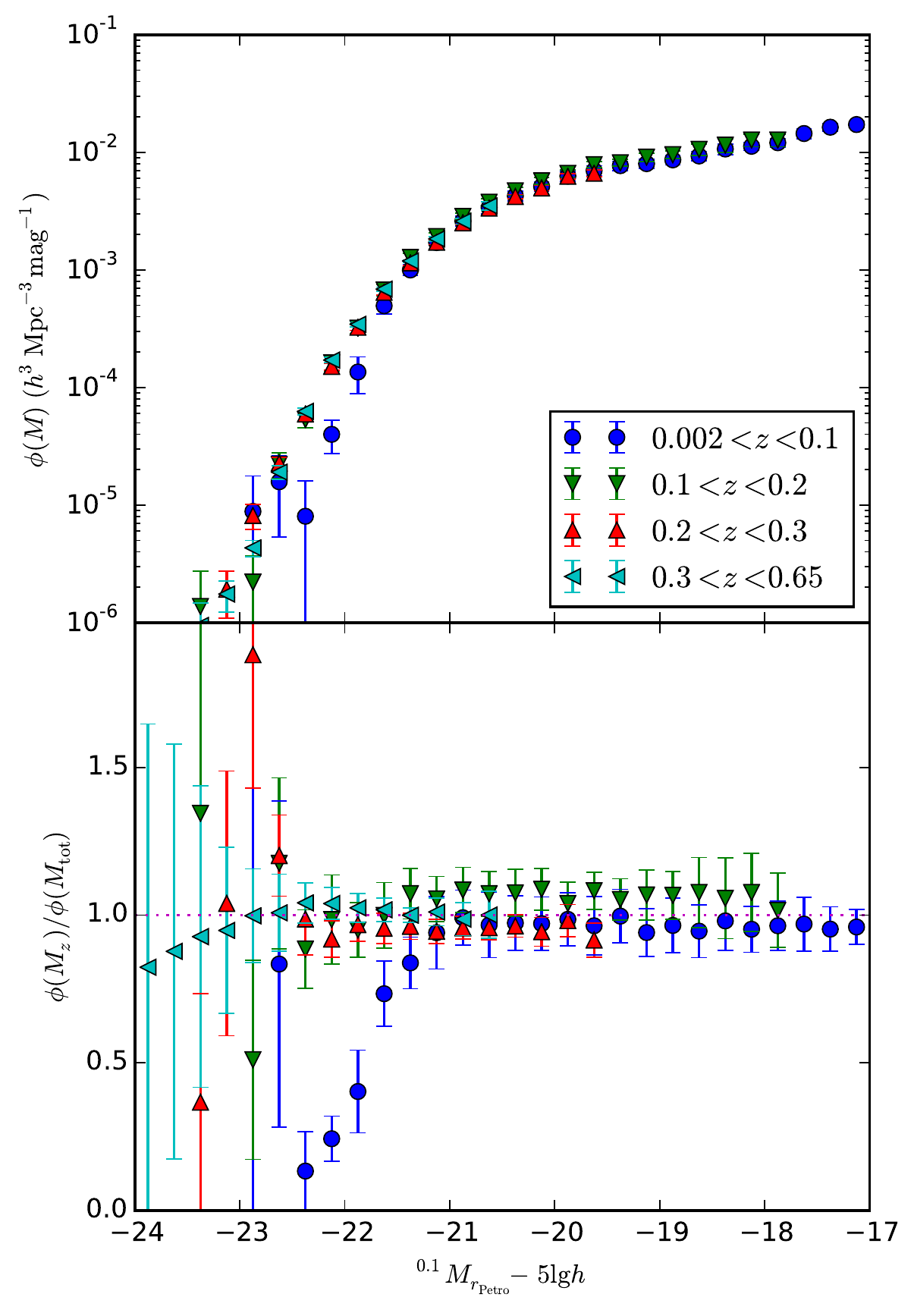}
\caption{Top panel: Petrosian $r$-band LFs measured in redshift slices 
as indicated for the full GAMA sample, 
applying evolutionary corrections as given in Table~\ref{tab:ev_fits}.
Bottom panel: The same LFs relative to the overall LF from the top panel
of Fig.~\ref{fig:lfr}.
}
\label{fig:lf_z_petro}
\end{figure}

\begin{figure}
\includegraphics[width=\linewidth]{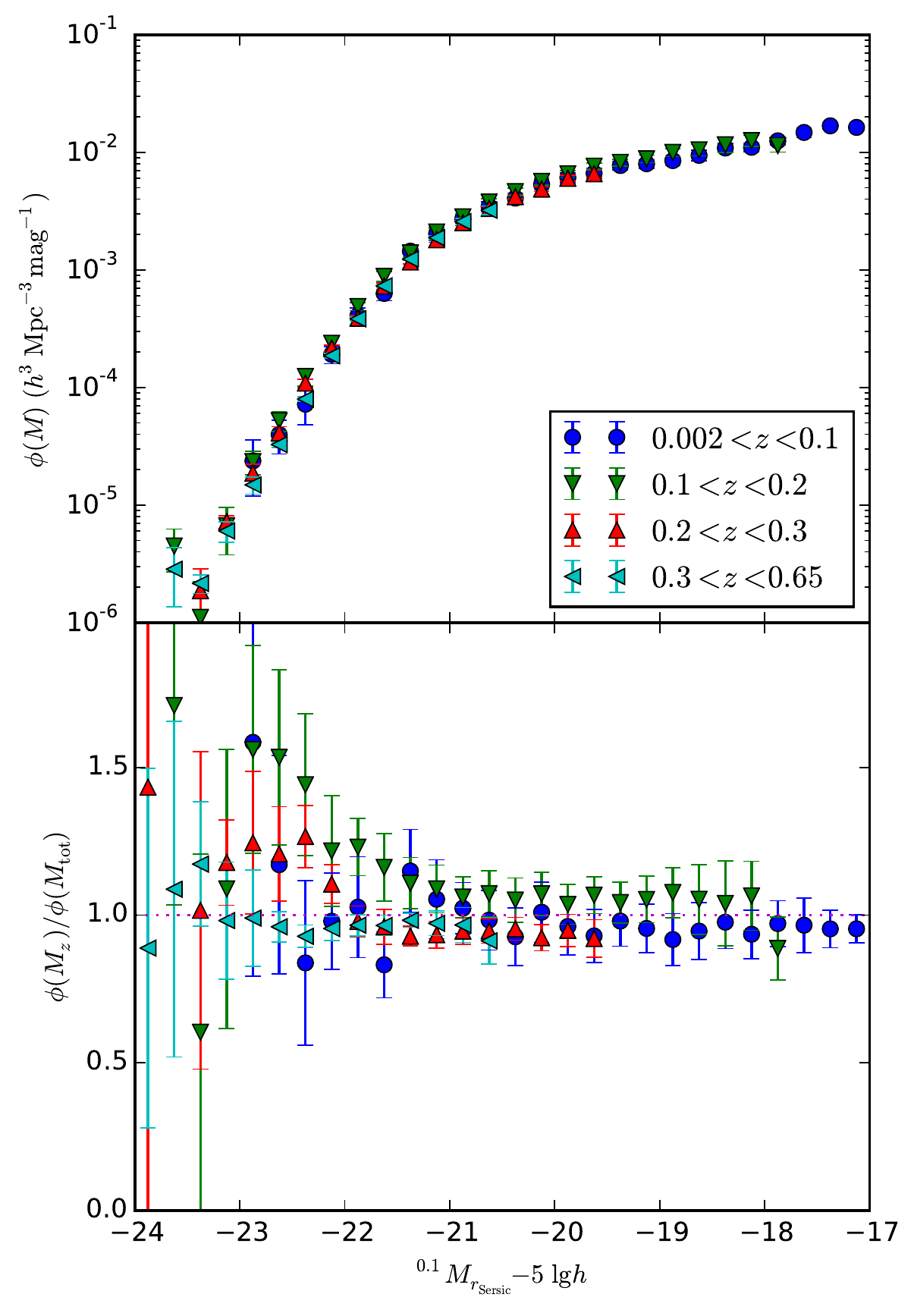}
\caption{As Fig.~\ref{fig:lf_z_petro} but using \Sersic\ magnitudes.
}
\label{fig:lf_z_sersic}
\end{figure}

In Fig.~\ref{fig:lf_z_petro}, we investigate how faithfully our simple evolution
model, namely one in which log-luminosity and log-density evolve linearly 
with redshift, is able to match the GAMA LF measured in redshift slices.
The top panel shows Petrosian $r$-band  LFs for the full GAMA-II sample 
measured in four redshift slices as indicated, 
calculated using equation~(\ref{eq:phi}) with the best-fit evolution parameters
and radial overdensities, and taking into account the appropriate
redshift limits.
If the evolution model accurately reflects true evolution,
and if we have successfully corrected for density variations,
then these LFs should be consistent where they overlap in luminosity.
In the bottom panel we have divided each LF by the LF determined from the
full sample ($0.002 < z < 0.65$; top panel of Fig.~\ref{fig:lfr}) 
in order to make differences more clearly visible.
We see that the lowest redshift LF, $z < 0.1$, is about 10--20 per cent lower 
than the $0.1 < z < 0.2$ LF, indicating that the linear evolution
model is somewhat undercorrecting at the lowest redshifts.
The low-redshift underdensity is particularly severe at the bright end:
the most luminous ($M_r - 5 \lg h \la -21.5$ mag) 
galaxies are underdense by $\sim 50$ per cent relative to the higher 
redshift slices.

In order to investigate these discrepancies further, we repeat this analysis
using redshift-sliced LFs determined using \Sersic\ magnitudes, with results
shown in Fig.~\ref{fig:lf_z_sersic}.
The underdensity of luminous, low-redshift galaxies is now much less severe;
instead we see an increased scatter between redshifts at the bright end,
with perhaps the $0.1 < z < 0.2$ LF biased high relative to the others.
It thus seems likely that the underdensity of luminous, 
low-redshift galaxies apparent in Fig.~\ref{fig:lf_z_petro} is largely due
to Petrosian magnitudes missing a significant fraction of the flux of
luminous galaxies, which will tend to have a de Vaucouleurs-like profile.
This problem is further exacerbated for such galaxies at low-redshift, 
which will have large angular extent, and thus also be susceptible 
to poor background subtraction: \citet{Blanton2011} show that galaxies 
of radius $\sim 100$ arcsec have their
magnitudes underestimated by around 1.5 mag in the SDSS DR7 database.
Both \Sersic\ and Petrosian photometry are subject to over-deblending
or `shredding' of large galaxy images.
When running the \Sersic\ fitting pipeline, 
\citet{Kelvin2012} aimed towards undershredding, as they were specifically
focussed on the primary galaxies in systems with close neighbours.
However, this does mean that the \Sersic\ fluxes
become susceptible to non-detection of nearby secondary sources,
which introduces a positive flux bias in crowded fields for a small fraction
of galaxies, see Section~\ref{sec:data}.

In conclusion, while the redshift-sliced Petrosian LFs do show some systematic
differences, use of GAMA-measured \Sersic\ magnitudes, 
which capture a larger fraction of total flux for de Vaucouleurs-profile 
galaxies, and which have an improved background subtraction compared with
SDSS DR7, largely mitigates these differences, and suggests that our
evolution model is a reasonable one.

\section{Discussion} \label{sec:discussion}

\subsection{Comprison with previous results}

While our evolution-corrected LFs agree well with previous estimates,
our finding of positive density evolution (in the sense
that comoving density was higher in the past) is at odds with most previous work
which has tended to find either mildly negative \citep{Cool2012}
or insignificant \citep{Blanton2003,Moustakas2013} density evolution.
\citet{Faber2007} find a declining comoving number density with redshift for
their red sample, with no noticeable density evolution for their blue
and full samples.
\citet{Zucca2009} also find a declining comoving number density with redshift 
for their reddest sample; for their bluest galaxies, they find increasing
number density with redshift.

At least some of the discrepancy between the sign of the density
evolution between us and e.g. \citet{Cool2012} might be explained 
by the way in which the LF and evolution are fitted.
\citet{Cool2012} fit the characteristic magnitude $M^*$
to each redshift range using the \citet{Sandage1979} maximum-likelihood method,
holding the faint-end slope parameter $\alpha$ fixed at its best-fit value
for the lowest-redshift range.
They then find the normalization $\phi^*$ using the \citet{Davis1982}
minimum-variance estimator.
Any over-estimate of luminosity evolution would lead to a
corresponding under-estimate in density evolution, due to 
the assumption of an unchanging faint-end slope with redshift and the
strong correlation between Schechter parameters.
Although any determination of evolution will be affected by degeneracies
between luminosity and density evolution, our method makes no assumption about
the (unobserved) faint-end slope of the LF at higher redshifts.
On the other hand, we do assume a parametric form for evolution.

It is also plausible that the discrepancies between estimated evolution 
parameters are due to the uncertainties in incompleteness-correction
required when analysing most galaxy surveys.
For example, when we cap our incompleteness-correction weights to 5, 
we see a reduction in the estimated density evolution parameters.
There are likely to be other effects leading to systematic errors 
in the determination of evolution parameters, 
which are not reflected in the (statistical) error contours.

A positive density evolution for the All galaxy sample would suggest 
a reduction in the number of galaxies with cosmic time, either through merging,
or due to galaxies dropping out of the sample selection criteria as they 
passively fade.
Neither scenario seems terribly likely;
\citet{Robotham2014} see evidence for only a small merger rate 
in the GAMA sample.
Perhaps a more likely explanation is that the apparent density evolution at low
redshift is actually caused by a local underdensity, e.g. 
\citet{Keenan2013,Whitbourn2014}.

\subsection{Future work}

\begin{figure}
\includegraphics[width=\linewidth]{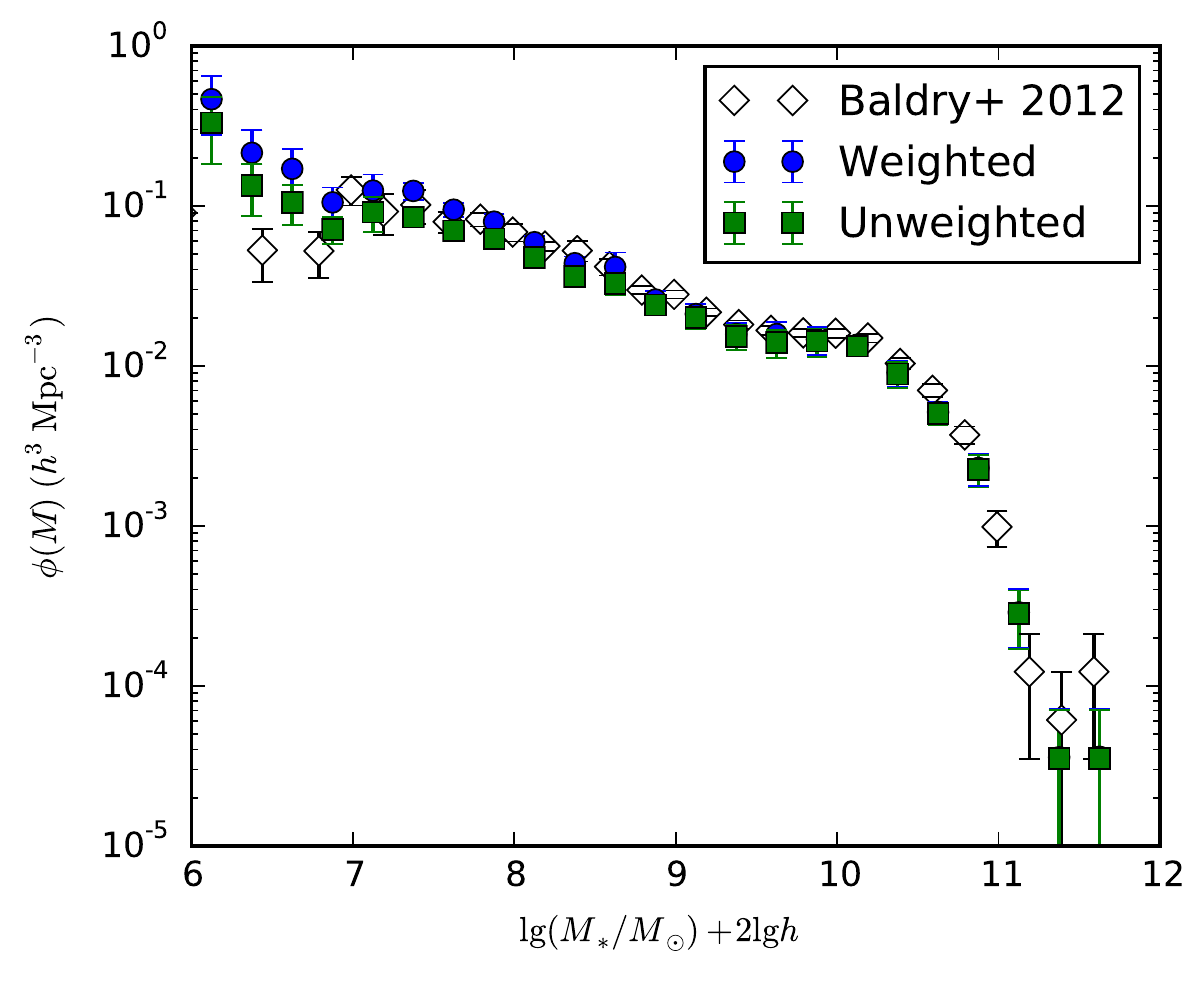}
\caption{Stellar mass function for low-redshift galaxies ($z < 0.06$)
determined from GAMA-II with (blue circles) and without (green squares)
applying a weighting correction for surface-brightness and redshift 
incompleteness.
Comparison data points for GAMA-I data from \citet{Baldry2012}
are shown as open diamonds.
}
\label{fig:smf_loz}
\end{figure}

There are several ways in which the present work can be extended.

Having derived density-corrected \Vmax\ values for each galaxy,
it is then trivial to determine other distribution functions,
such as the stellar mass and size functions, and their evolution.
By way of a quick example, in Fig.~\ref{fig:smf_loz} we plot the
stellar mass function for low-redshift ($z < 0.06$) GAMA-II galaxies, 
using the stellar mass estimates of \citet{Taylor2011}.
In the mass regime where surface-brightness completeness is high,
$\log(M/M_\odot) + 2 \log h \ga 8$, we find excellent agreement with the
earlier estimate from \citet{Baldry2012}
using a density-defining population.
The upturn seen in the mass function below 
$\log(M/M_\odot) + 2 \log h \approx 7$
will be sensitive to the incompleteness corrections applied;
confirmation of this feature will need to await the availability of deeper
VLT Survey Telescope Kilo-degree Survey (VST KiDS) imaging in the GAMA regions.
Future work will explore the evolution of the stellar mass function.

The density-corrected \Vmax\ values will also be used to generate 
the radial distributions of random points
required to measure the clustering of flux-limited galaxy samples 
(Farrow et al., in prep.)

We plan to explore 
the possibility of using 
the \citet{Taylor2011} 
stellar population synthesis (SPS)
model fits to GAMA data to derive luminosity evolution parameters
$Q_e$ for individual galaxies.
If the models can predict $Q_e$ with sufficient reliability,
the degeneracy in fitting for both luminosity
and density evolution would be largely eliminated.
This would also allow for the fact that galaxies
have individual evolutionary histories.

We also plan to incorporate the environmental-dependence of the LF into 
our model.
Note that the radial overdensities shown in
Fig.~\ref{fig:delta} are a poor estimate of the density around each galaxy
since they are averages over the entire GAMA-II area within each redshift shell.
\citet{McNaught-Roberts2014} present estimates of the LF
for galaxies in bins of density within $8 \hMpc$ spheres.
We are currently extending density estimation to the full GAMA sample using
a variety of density measures (Martindale et al., in prep).

This main focus of this paper has been to correct the LF and radial density 
for the effects of evolution, rather than to measure evolution per se.
An alternative way of constraining evolution is to  measure how the 
luminosity of galaxies at a fixed space density evolves. 
Via comparison with a model for the evolution of stellar populations 
(or luminosity evolution of the fundamental plane), 
one can estimate the rate of mass growth, e.g. \citet{Brown2007}.

\section{Conclusions} \label{sec:concs}

We have described an implementation of the \citet{Cole2011} JSWML method used to
infer the evolutionary parameters, the radial density variations 
and the $r$ band LF of galaxies in the GAMA-II survey.
For the overall population, we find that galaxies have faded in $r$-band
luminosity by about 0.5 mag, and have decreased in comoving number density 
by a factor of about 1.6 since $z \approx 0.5$, i.e. over the last 5~Gyr or so.
When the population is divided into red and blue galaxies, the differences 
in density evolution parameter $P_e$ are statistically insignificant.
Luminosity evolution is significantly stronger for blue galaxies than for red.
Evolution in the luminosity density evolution of blue galaxies is higher than
that of red at the $\sim 5$-$\sigma$ level.
These findings are consistent with those of \citet{Loveday2012} based
on GAMA-I and are as expected, since a fraction of galaxies that were 
blue in the past will have since ceased star formation and become red.

While there still exists 
some
degeneracy between
the parameters describing luminosity ($Q_e$) and density ($P_e$) evolution
for GAMA-II data, 
see Fig.~\ref{fig:chigrid},
analysis of simulated data and comparison with a local
galaxy sample from SDSS \citep{Blanton2005} shows that we are able to recover
the evolution-corrected LF to high accuracy.
In detail, GAMA LFs are poorly described by Schechter functions, 
due to excess number density at both faint and bright luminosities,
particularly for the red population.

The density-corrected \Vmax\ values will be made available via the GAMA 
database.

\section*{Acknowledgements}

JL acknowledges support from the Science and Technology Facilities Council
(grant number ST/I000976/1) and illuminating discussions with Shaun Cole.
PN acknowledges the support of the Royal Society through the award
of a University Research Fellowship, the European Research
Council, through receipt of a Starting Grant (DEGAS-259586)
and the Science and Technology Facilities Council (ST/L00075X/1).
It is also a pleasure to thank the referee, Thomas Jarrett,
for his careful reading of the
manuscript and for his many useful suggestions.

GAMA is a joint European-Australasian project based around a
spectroscopic campaign using the Anglo-Australian Telescope. The GAMA
input catalogue is based on data taken from the Sloan Digital Sky
Survey and the UKIRT Infrared Deep Sky Survey. Complementary imaging
of the GAMA regions is being obtained by a number of independent
survey programs including GALEX MIS, VST KIDS, VISTA VIKING, WISE,
Herschel-ATLAS, GMRT and ASKAP providing UV to radio coverage. GAMA is
funded by the STFC (UK), the ARC (Australia), the AAO, and the
participating institutions. The GAMA website is:
\url{http://www.gama-survey.org/}.

\bibliographystyle{mn2e}
\setlength{\bibhang}{2.0em}
\setlength\labelwidth{0.0em}
\bibliography{library}{}

\end{document}

%% file: authors.tex
\author[J.~Loveday et al.]
{{\parbox{\textwidth}{\raggedright J.~Loveday,$^{1}$\thanks{E-mail:~J.Loveday@sussex.ac.uk}
P.~Norberg,$^{2}$
I.K.~Baldry,$^{3}$
J.~Bland-Hawthorn,$^{4}$
S.~Brough,$^{5}$
M.J.I.~Brown,$^{6}$
S.P.~Driver,$^{7,8}$
L.S.~Kelvin,$^{9}$
S.~Phillipps$^{10}$
}}\\
\vspace{0.4cm}\\
{\parbox{\textwidth}{\raggedright 
$^{1}$Astronomy Centre, University of Sussex, Falmer, Brighton BN1 9QH
\\
$^{2}$Institute for Computational Cosmology, Department of Physics,
Durham University, South Road, Durham DH1 3LE
\\
$^{3}$Astrophysics Research Institute, Liverpool John Moores University, IC2, Liverpool Science Park, 146 Brownlow Hill, Liverpool, L3 5RF
\\
$^{4}$Sydney Institute for Astronomy, School of Physics, University of
Sydney, NSW 2006, Australia
\\
$^{5}$Australian Astronomical Observatory, PO Box 915, North Ryde, NSW 1670, Australia
\\
$^{6}$School of Physics, Monash University, Clayton, Victoria 3800, Australia
\\
$^{7}$International Centre for Radio Astronomy Research (ICRAR),
    The University of Western Australia, 35 Stirling Highway, Crawley,
    WA6009, Australia
\\
$^{8}$School of Physics \& Astronomy, University of St Andrews, North
Haugh, St Andrews, KY16 9SS
\\
$^{9}$Institut f\"{u}r Astro- und Teilchenphysik, Universit\"{a}t Innsbruck, Technikerstra{\ss}e 25, 6020 Innsbruck, Austria
\\
$^{10}$Astrophysics Group, HH Wills Physics Laboratory,
University of Bristol, Tyndall Avenue, Bristol BS8 1TL
\\
}}}

%% file: ev_table.tex
    \begin{math}
    \begin{array}{lcccccc}
    \hline
    {\rm Sample} & Q_e & P_e & Q_e + P_e & m & c & \chi^2_\nu \\
    
        \hline
        \multicolumn{7}{c}{\mbox{Mean Probability}} \\
        \hline
        All & 1.03 \pm 0.10 & 1.00 \pm 0.25 & 2.02 \pm 0.05 & -0.36 & 1.38 & \ldots \\
                Blue & 1.09 \pm 0.10 & 1.30 \pm 0.25 & 2.39 \pm 0.04 & -0.35 & 1.55 & \ldots \\
                Red & 0.58 \pm 0.18 & 1.55 \pm 0.40 & 2.12 \pm 0.08 & -0.38 & 1.17 & \ldots \\
                
        \hline
        \multicolumn{7}{c}{\mbox{LF-redshift}} \\
        \hline
        All & 1.03 \pm 0.07 & 1.00 \pm 0.20 & 2.02 \pm 0.05 & -0.35 & 1.37 & 3.76 \\
                Blue & 1.18 \pm 0.05 & 1.07 \pm 0.15 & 2.25 \pm 0.04 & -0.34 & 1.55 & 3.46 \\
                Red & 0.73 \pm 0.10 & 1.25 \pm 0.25 & 1.98 \pm 0.06 & -0.36 & 1.16 & 3.35 \\
                
    \hline
    \end{array}
    \end{math}
    

%% file: lf_table.tex
    \begin{math}
    \begin{array}{lcccc}
    \hline
    {\rm Sample} & \alpha & M^* - 5 \log h & \log \phi^*/\denunit & \chi^2_\nu\\
    
        \hline
        \multicolumn{5}{c}{\mbox{Mean Probability Petrosian}} \\
        \hline
        All & -1.26 \pm 0.07 & -20.71 \pm 0.05 & -2.02 \pm 0.04 & 2.33\\
                Blue & -1.38 \pm 0.06 & -20.36 \pm 0.05 & -2.27 \pm 0.05 & 0.96\\
                Red & -0.79 \pm 0.11 & -20.68 \pm 0.06 & -2.23 \pm 0.05 & 3.48\\
                
        \hline
        \multicolumn{5}{c}{\mbox{LF-redshift Petrosian}} \\
        \hline
        All & -1.26 \pm 0.07 & -20.71 \pm 0.05 & -2.02 \pm 0.04 & 2.33\\
                Blue & -1.37 \pm 0.06 & -20.35 \pm 0.05 & -2.24 \pm 0.05 & 1.01\\
                Red & -0.77 \pm 0.11 & -20.64 \pm 0.05 & -2.20 \pm 0.04 & 3.01\\
                
        \hline
        \multicolumn{5}{c}{\mbox{LF-redshift Sersic}} \\
        \hline
        All & -1.30 \pm 0.06 & -20.88 \pm 0.06 & -2.13 \pm 0.04 & 2.96\\
                Blue & -1.39 \pm 0.07 & -20.40 \pm 0.06 & -2.27 \pm 0.05 & 1.03\\
                Red & -0.79 \pm 0.12 & -20.72 \pm 0.07 & -2.21 \pm 0.05 & 4.91\\
                
    \hline
    \end{array}
    \end{math}
    